\documentstyle[11pt,aaspp4]{article}
\lefthead{I. V. Igumenshchev and M. A. Abramowicz}
\righthead{Two-dimensional accretion flows}

\begin{document}

\title{Two-dimensional models of hydrodynamical
       accretion flows into black holes}

\author{Igor V. Igumenshchev\altaffilmark{1} and
        Marek A. Abramowicz\altaffilmark{2}}
\affil{Institute of Theoretical Physics, G\"oteborg University and Chalmers
University of Technology, 412~96 G\"oteborg, Sweden;
ivi@fy.chalmers.se, marek@fy.chalmers.se}

\altaffiltext{1}{Institute of Astronomy,
48 Pyatnitskaya St., 109017 Moscow, Russia}
\altaffiltext{2}{Laboratorio Interdisciplanare SISSA and ICTP, Trieste, Italy}

\begin{abstract}

We present a systematic numerical study of 
two-dimensional axisymmetric accretion flows around black holes.
The flows have no radiative cooling and are treated in the 
framework of the hydrodynamical approximation. 
The models calculated in this study cover the large range of the 
relevant parameter space.
There are four types of flows, determined by the values of the viscosity
parameter $\alpha$ and the adiabatic index $\gamma$: convective flows,
large-scale circulations, pure inflows and bipolar outflows. 
Thermal conduction introduces
significant changes to the solutions, but does not create a new flow type.
Convective accretion flows and flows with large-scale circulations
have significant outward-directed energy fluxes,
which have important implications for the spectra and luminosities
of accreting black holes.

\noindent {\em Subject headings:} Accretion, accretion disks ---
conduction --- convection --- hydrodynamics --- turbulence

%\keywords{accretion, accretion disks --- convection --- hydrodynamics 
%--- turbulence}

\end{abstract}

\section{Introduction}

In this paper we present the 
%most complete and up to date 
numerical study
of global properties of hydrodynamical black hole accretion flows
with a very inefficient radiative cooling. Such flows, coined ADAFs by
Lasota (1996, 1999), are thought to be present in several astrophysical
black hole candidates, in particular in low mass X-ray binaries 
and in some active
galactic nuclei. Observed properties of ADAFs may be directly
connected to black hole physics, and for this reason ADAFs have recently
attracted a considerable attention 
%of both experts and general public 
(for reviews see e.g. Kato, Fukue \& Mineshige 1998; 
Abramowicz, Bj\"ornsson \& Pringle 1998; Narayan 1999).

The radiation losses are unimportant for dynamics as well as
for thermal balance of
ADAFs, and therefore the details of radiative processes are not crucial.
Radiative feed-back into hydrodynamics is 
%in most cases
negligible and may be treated as a small perturbation\footnote{Objects
that are {\it external} to ADAFs may change the balance: for example, if
there is an external source of soft photons, they may provide an
additional, possibly efficient, Compton cooling of optically thin plasma
(Shapiro, Lightman \& Eardley 1976).}. 
Abramowicz et al. (1995) and in more details Chen et al. (1995) described
accretion disks solutions in the parameter space 
($\dot{m}$, $\tau$, $\alpha$), where
$\dot{m}=\dot{M}/\dot{M}_{Edd}$ 
is the accretion rate expressed in the Eddington units, $\tau$ is the
optical depth and $\alpha$ is the viscosity parameter. 
In this space ADAFs exist in two regimes
(as anticipated by Rees et al. 1982):

\noindent
(1) ADAFs with $\tau\gg 1$ have super-Eddington accretion rates, 
$\dot{m} > 1$. Radiation is trapped inside the accretion flow
(Katz 1977; Begelman 1978). To this category belong
slim accretion disks (Abramowicz et al. 1988) which have
a vertical scale comparable the corresponding radius.

\noindent
(2) ADAFs with $\tau\ll 1$ have very sub-Eddington accretion rates, 
$\dot{m}\ll 1$. 
These flows have been first investigated by Ichimaru (1977), but
the recent interest in them was generated mostly by the works of Narayan
and his collaborators, after important aspects of nature of the flows
had been explained by
Narayan \& Yi (1994, 1995b) and Abramowicz et al. (1995).

The most important parameters for the physics of ADAFs are 
the viscosity parameter $\alpha$ and the adiabatic index $\gamma$.
The latter parameter determines the regime of ADAFs through 
the equation of state. 
The parameters $\tau$ and $\dot{m}$ are not important: ADAFs have either
$\tau\gg 1$ and $\dot{m}> 1$, or $\tau\ll 1$ and $\dot{m}\ll 1$.
There are no strong observational or theoretical limits for 
$\alpha$ and $\gamma$, and therefore, at present, 
one needs to construct models in wide ranges of them. 
In this paper we have constructed models for
$10^{-2}\le\alpha\le 1$ and $\gamma = 4/3$, $3/2$, $5/3$.
Our models are time dependent and fully 2-D: all
components of forces and all components of viscous stresses are included
in the calculations. 
To minimize the influence of the outer boundary condition onto 
the flow structure, we consider the solutions in the large radial range,
$3r_g \le r \le 8\times 10^3r_g$, where $r_g = 2GM/c^2$ is the gravitational 
radius of the central black hole with mass $M$.
We have found that the properties of ADAFs depend mainly on the viscosity, 
i.e. on $\alpha$, and also, but less strongly, on the adiabatic
index $\gamma$. Four types of accretion flows can be distinguished
(see Figure 1). 

\noindent
(i) Convective flows. 
For a very small viscosity, $\alpha\la 0.03$,
ADAFs are convectively unstable,
as predicted by Narayan \& Yi (1994) and confirmed in numerical simulations
by Igumenshchev, Chen \& Abramowicz (1996), Stone, Pringle \& Begelman (1999)
and Igumenshchev \& Abramowicz (1999).
Axially symmetric convection transports the
angular momentum {\it inward} rather than outward, 
for a similar reason
as that described by Stone \& Balbus (1996) in the context of turbulence: 
if there are no azimuthal gradients of pressure, 
turbulence tries to erase the angular momentum gradient.
This property of convection governs the flow structure, as it was shown
for ADAFs by Narayan, Igumenshchev \& Abramowicz (2000) 
and Quataert \& Gruzinov (2000).
There are small scale circulations, with the matter fluxes
considerably greater than that entering the black hole.
Convection transports a significant amount of 
the dissipated binding energy outward. No powerful outflows are present.

\noindent
(ii) Large-scale circulations. 
For a larger, but still small viscosity, $\alpha\sim 0.1$,
ADAFs could be both stable or unstable convectively,
depending on $\alpha$ and $\gamma$. The flow pattern consists of
the large-scale ($\sim r$) meridional circulations.
No powerful unbound outflows are present.
In some respect this type of flow is the limiting case of the convective
flows in which the small scale motions are suppressed by larger viscosity.

\noindent
(iii) Pure inflows. 
With an increasing viscosity, $\alpha\simeq 0.3$,
the convective instability dies off.
Some ADAFs (with $\gamma\simeq 3/2$)
are characterized by a pure inflow pattern, and agree in many
aspects with the self-similar models (Gilham 1981; Narayan \& Yi 1994). 
No outflows are present.

\noindent
(iv) Bipolar outflows. 
For a large viscosity, $\alpha\simeq 1$, 
ADAFs differ considerably from the simple
self-similar models. Powerful unbound bipolar outflows are present.

Effects of turbulent thermal
conduction have been studied in several simulations.
The conduction has an important influence to the flow structure, 
but it does not introduce a new type of flow.

The paper is organized as follows.
In \S2 we describe equations, numerical method and
boundary conditions. In \S3 we present numerical results for
models with and without thermal conduction. In \S4 we discuss
the properties of the solutions and their implications.
In \S5 we give the final conclusions.

\section{Numerical method}

We compute ADAFs models by solving the
non-relativistic time-dependent Navier-Stokes equations
%with inclusion of thermal conduction
which describe accretion flows in a given and fixed gravitational field:
$$
{d \rho\over dt}+\rho\nabla\vec{v}=0, \eqno(2.1)
$$
$$
\rho{d \vec{v}\over dt}=-\nabla P + \rho\nabla\Phi + \nabla{\bf \Pi},
\eqno(2.2)
$$
$$
\rho{d e \over dt}=-P\nabla\vec{v}-\nabla\vec{q}+Q. \eqno(2.3)
$$
Here $\rho$ is the density, $\vec{v}$ is the velocity,
$P$ is the pressure,
$\Phi=-GM/r$ is the Newtonian gravitational potential of the central point
mass $M$,
$e$ is the specific thermal energy, ${\bf \Pi}$ is the
viscous stress tensor with all components included,
$\vec{q}$ is the heat flux density due to thermal conduction
and $Q$ is the dissipation function.
The flow is assumed to be axially symmetric.
There is no radiative cooling of the accretion gas.
We adopt the ideal gas equation of state, 
$$
P=(\gamma-1)\rho e, \eqno(2.4)
$$
%where $\gamma$ is the adiabatic index,
and consider only the shear viscosity with the kinematic viscosity
coefficient given by
$$
\nu=\alpha{c_s^2\over \Omega_K}, \eqno(2.5)
$$
where $0<\alpha\la 1$,
$c_s=\sqrt{P/\rho}$ is the isothermal sonic speed and
$\Omega_K=\sqrt{GM/r^3}$ is the Keplerian angular velocity.
We assume that the thermal conduction heat flux is directed down the specific
entropy gradient,
$$
\vec{q}=-\chi\rho T\nabla s, \eqno(2.6)
$$
where $s$ is the specific entropy, $T$ is the temperature
and $\chi$ is the thermometric conductivity.
The formula (2.6) is
correct for flows in which the heat conduction is due to
either turbulent eddies or diffusion of radiation 
in the optically thick medium.
Other laws of thermal conduction
are possible for different heat transfer mechanisms.
In numerical models we assume a simplified dependence,
$$
\chi=\nu/Pr, \eqno(2.7)
$$ 
where $Pr$ is the dimensionless
Prandtl number assumed to be constant in the flow, 
and $\nu$ is defined by (2.5). In the flows without thermal conduction,
i.e. $\chi=0$, one formally has $Pr=\infty$. 
In the case of turbulent flows the actual value of the Prandtl number is not
clearly known and can vary in a wide range,
$1\la Pr < \infty$, depending on the nature of turbulence.
If the viscosity in turbulent flows is provided mainly by small scale eddies,
one can expect $Pr\sim 1$. If the viscosity is due to magnetic
stress, the thermal conduction could be significantly suppressed, 
i.e. $Pr\gg 1$.
Note, that for molecular thermal conduction
in gases, the Prandtl number
is always of the order of unity (e.g. Landau \& Lifshitz 1987).
In actual calculations, we use 
$$
\vec{q}=-\chi\left[\nabla(\rho e)-\gamma e \nabla\rho\right], \eqno(2.8)
$$
which is equivalent to (2.6).

We split the numerical integration of equations (2.1)-(2.3)
into three sub-steps:
hydrodynamical, viscous and conductive. 
The hydrodynamical sub-step is calculated by using
the explicit Eulerian finite-difference algorithm PPM developed by
Colella \& Woodward (1984). The viscous sub-step is
solved by applying an implicit method with a direction-splitting
when calculating the contributions to equation (2.2).
The dissipation function $Q$ in equation (2.3) 
is calculated explicitly.
The conductive sub-step in equation (2.3) is again
calculated by an implicit method with a direction-splitting.
The time-step $\Delta t$ for the numerical integration
is chosen in accordance with the Courant
condition for the hydrodynamical sub-step.

We use a spherical grid
$N_r\times N_\theta=130\times 50$ with the inner radius at $r_{in}=3 r_g$ and
the outer radius at $r_{out}=8000 r_g$.
The grid points are logarithmically spaced in the radial direction and
uniformly spaced in the polar direction from $0$ to $\pi$.
The general-relativistic capture effect, that governs the flow
near black hole, is modeled by using an absorbing
boundary condition at $r=r_{in}$.
We assume no viscous angular momentum and energy fluxes
through the inner boundary associated with the ($r\theta$) and
($r\phi$) components of the shear stress. Namely, we assume
$d(v_\theta/r)/dr=0$ and $d(v_\phi/r)/dr=0$ at $r=r_{in}$.
Also, the energy flux due to the thermal conduction at $r=r_{in}$
is assumed to be zero.
At the outer boundary $r_{out}$ we apply an absorbing boundary condition.
The matter can freely outflow through $r_{out}$, but there are no
flows from outside.

In the calculations, we assume that mass is steadily injected into
the calculation domain
from an equatorial torus near the outer boundary of the grid.
Matter is injected there with angular momentum equal to
$0.95$ times the Keplerian angular momentum.
Due to viscous spread, a part of the injected matter moves inwards
and forms an accretion flow. The other part leaves
the computation domain freely through the outer boundary.
We start computations from an initial state in which
there is a little amount of mass in the grid. 
As the injected mass spreads and accretes,
the amount of mass within the grid increases. 
After a time comparable
to the viscous time scale at $r_{out}$, the accretion flow achieves
a quasi-stationary behaviour, and may be considered to be in a steady state.
However, for some models this `steady state' is steady only in the sense of
time-averaging: these flows show persistent 
chaotic fluctuations, at any given point, which do not die out with time.

\section{Two-dimension hydrodynamical models}

Spatial re-scaling $r\rightarrow x$ together with time re-scaling
$t\rightarrow t/(r_g/c)$, make solutions of (2.1)--(2.3) independent of
the black hole mass $M$. There is also an obvious re-scaling of density
$\rho\rightarrow \rho/\dot{M}_{inj}$ that makes the solutions independent of
the mass injection rate $\dot{M}_{inj}$.
Thus, with fixed boundary conditions, the numerical models
are described by the three dimensionless parameters 
$\alpha$, $\gamma$ and $Pr$.
We have calculated a variety of models for different 
values of these parameters (see Table~1).

\placetable{tbl-1}

\subsection{Models without thermal conduction}

First, we describe accretion flows with no thermal conduction
($Pr=\infty$). The results of the simulations are summarized 
in Figure~1 (left panel).
Here circles show the location of the computed models
in the ($\alpha$, $\gamma$)
plane. Empty circles correspond to stable laminar flows.
Unstable models with large-scale ($\sim r$) circulation motions 
%with spatial scale $\sim r$,
are represented by
crossed circles. Unstable models with small-scale ($\la r$)
convective motions are
shown by solid circles.
Arrows indicate the presence of powerful
outflows or strong inflows in the models. Two outward directed arrows, on the
upper and lower of a circle, mean the bipolar outflows, whereas one arrow
corresponds to the unipolar outflow. 
Inward directed arrows indicate that the model has a pure inflow pattern.
Models which are not marked by arrows reveal neither powerful outflows,
nor pure inflow.

\begin{figure}
\plottwo{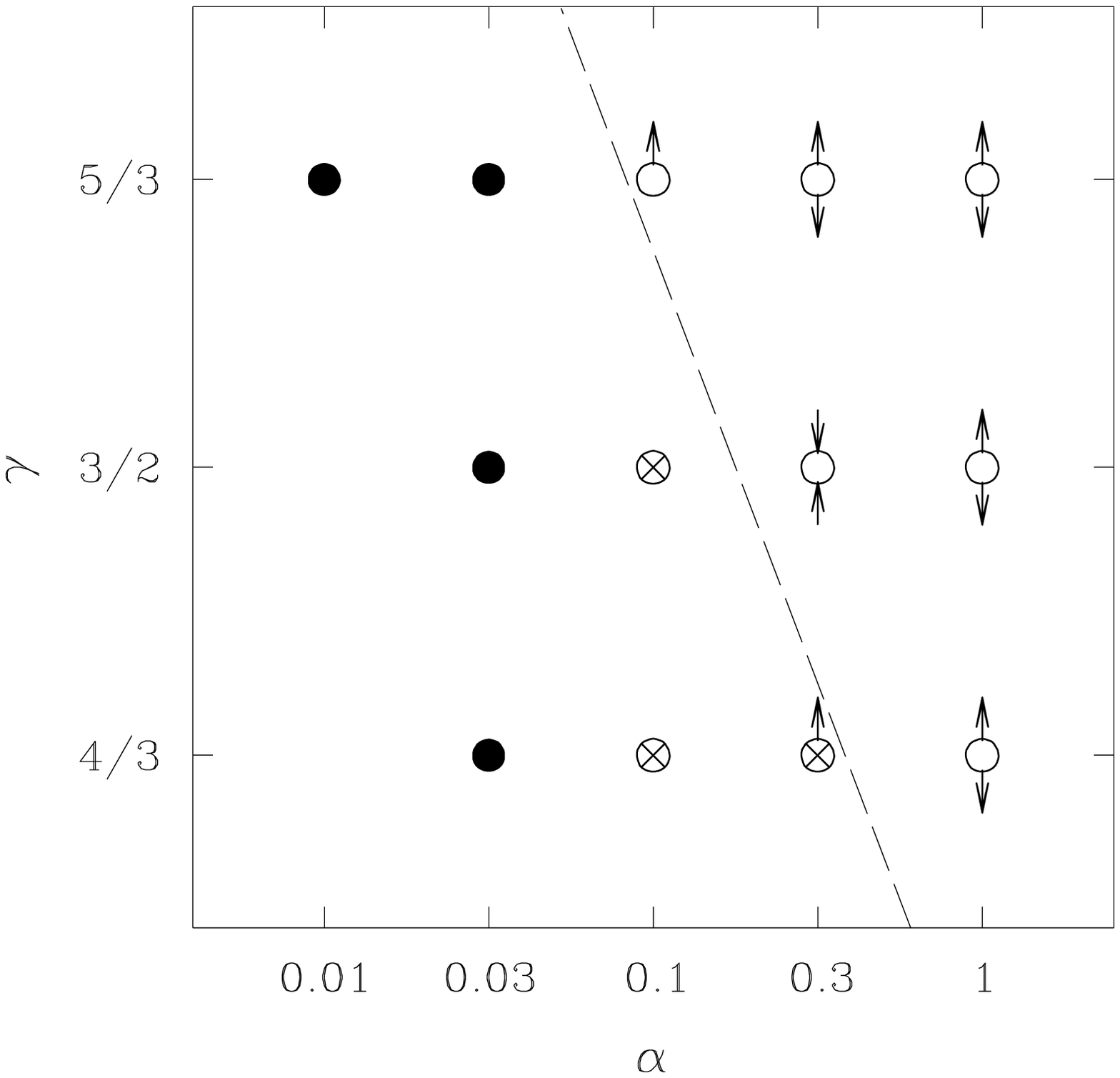}{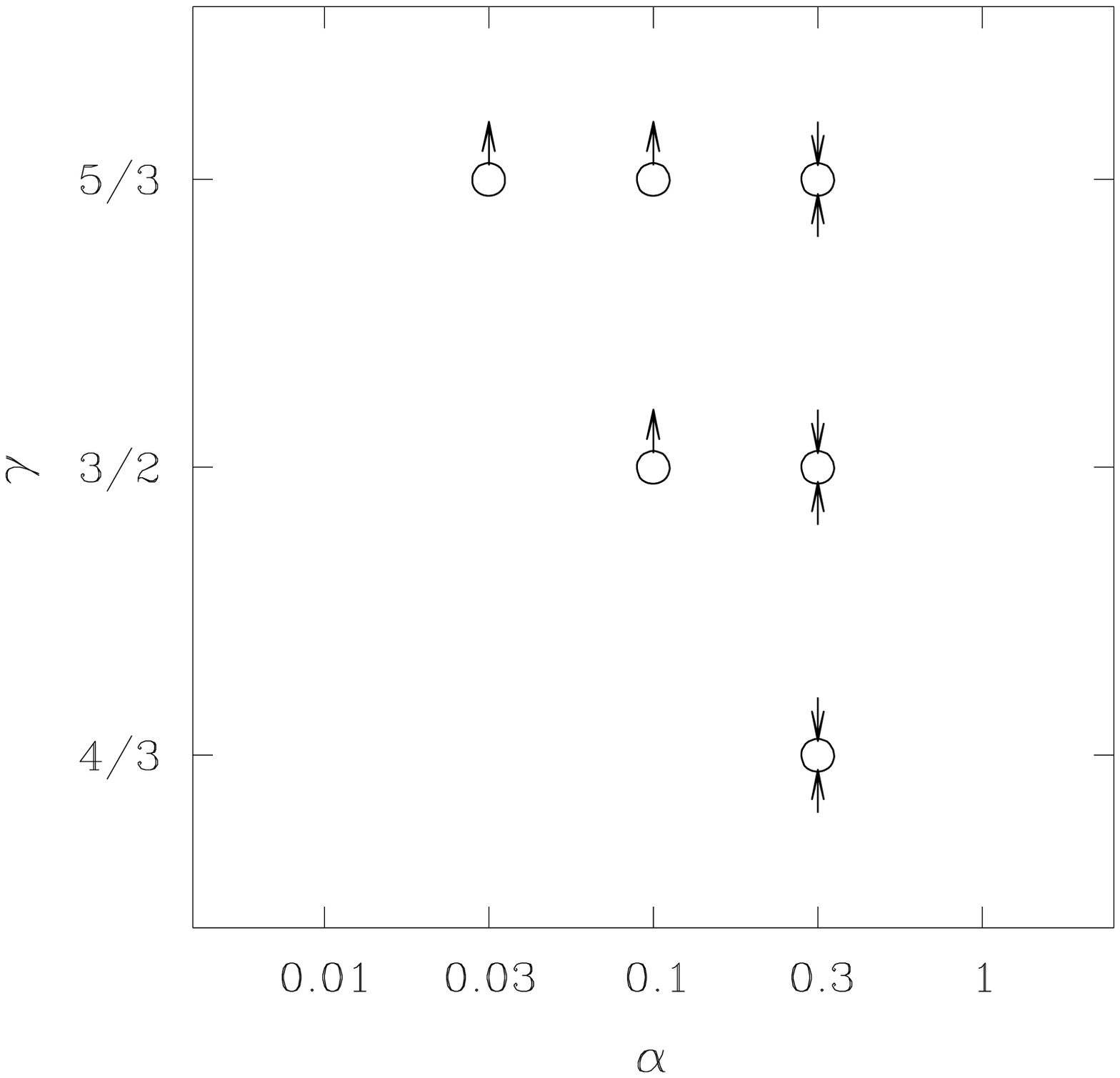}
\caption{Properties of two-dimensional axisymmetric models of accretion
flows. Models with no thermal conduction ($Pr=\infty$) are shown
in the left panel. Models with thermal conduction ($Pr=1$) are shown
in the right panel.
Each circle represents a model in the ($\alpha$, $\gamma$)
parameter space. The empty circles correspond to laminar flows,
the crossed circles represent unstable models with large-scale ($\sim r$)
meridional circulations of matter and solid circles indicate models 
with small-scale ($<r$) convective motions.
The arrows indicate powerful outflows or strong inflows in the models.
Two outward directed arrows correspond to bipolar outflows,
whereas one arrow corresponds to a unipolar outflow.
Two inward directed arrows correspond to the models with a pure inflow.
The models with bipolar outflows and pure inflow are highly symmetrical 
with respect
to the equatorial plane. The dashed line on the left panel
approximately separates regions of convectively stable/unstable flows.
\label{fig1}}
\end{figure}

\placefigure{fig1}

The source of matter in our models has a constant injection rate 
and locates close to the outer boundary. In a steady state one expects that
due to this particular location of the source,
most of the material injected into the computational domain escapes
through the outer boundary and only a minor part of it accretes 
into the black hole.
Table~2 presents the ratio of the mass accretion rate $\dot{M}_{0}$,
measured at the inner boundary, to the mass outflow 
rate $\dot{M}_{out}$ through the outer boundary for a variety of
models. The selected models are both stable and unstable, and 
we use the time-averaged rates in the case of unstable models. 
From Table~2 one can see that in most cases the ratio
$\dot{M}_{0}/\dot{M}_{out}$ is very small, $\sim 10^{-2}-10^{-3}$,
and shows a complicated dependence on $\alpha$ and $\gamma$.
The smallest relative accretion rates correspond to the models with a small 
viscosity (Models~J, K and L)
and a smaller ratio corresponds to a larger $\gamma$ in these models.
A similar dependence of $\dot{M}_{0}/\dot{M}_{out}$ on $\gamma$
can also be seen for other fixed values of $\alpha$.
However, Model~E demonstrates a peculiar property: the mass accretion rate
is about two times larger than the outflow rate.
We will discuss later in detail this peculiar model and show that it is 
closely related to the `standard' self-similar ADAF solutions 
(Narayan \& Yi 1994), 
whereas other models are either not related to self-similar ADAFs, or
related to self-similar ADAFs of a new kind (Narayan et al. 2000; Quataert \&
Gruzinov 2000).

\placetable{tbl-2}

A small value of the ratio $\dot{M}_{0}/\dot{M}_{out}$ does not 
indicate powerful unbound outflows. 
In some models the large outflow rate
is only a consequence of our choice of
the geometry of matter injection. To confirm this point,
we present in Figure~2 a variety of
histograms, which show the fraction of matter that outflows behind the outer
boundary and has a fixed value of the dimensionless Bernoulli parameter $Be$,
$$
Be=\left({1\over 2}v^2+W-{GM\over r}\right)
\left/{GM\over r}\right.. \eqno(3.1)
$$
Here $W=\gamma c_s^2/(\gamma-1)$ is the specific enthalpy.
The histograms are shown for Models~A, D, G and J, all with fixed $\gamma=5/3$ 
and different values of $\alpha$.
We have used the standard normalization
$\sum\Delta\dot{m}_{out}(Be)=1$ in the histograms.
The matter with positive $Be$ is gravitationally unbound and can form 
%(but does not have to have) 
outflows,
whereas the matter with negative $Be$ is gravitationally bound
and cannot escape to a large radial distance.
One can see in Figure~2 that the high viscosity flows (Models~A and D)
form powerful unbound outflows with positive and large $Be$ on average,
and only a minor part of the outflowing matter ($\sim 20\%$)
is gravitationally bound.
In the low viscosity flows (Models~G and J)  
most of the matter that moved through the
outer boundary has $Be<0$ and thus it remains
gravitationally bound and cannot form powerful outflows.
We have found a similar situation with respect to formation of 
bound/unbound outflows in models with different $\gamma$.
In general, flows with low $\alpha$ are bound and have no powerful outflows.
%This fully confirms the analytic result of 
%Abramowicz, Lasota \& Igumenshchev (2000).

\begin{figure}
\plotone{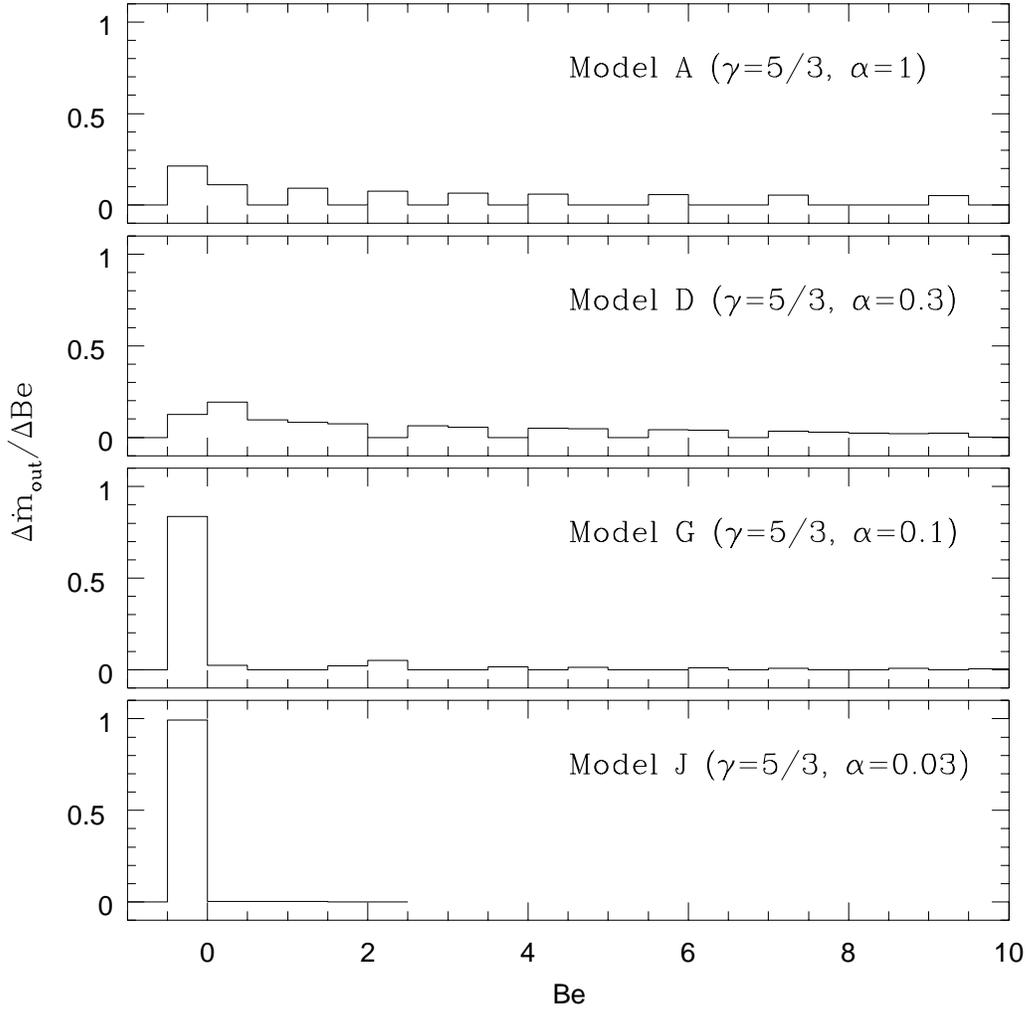}
\caption{Histograms showing the distribution of mass over the dimensionless
Bernoulli parameter $Be$ in the outflowing material 
measured at the outer boundary $r_{out}$. 
In the case of the large viscosity $\alpha\ga 0.3$ flows (Models~A and D)
most of the escaped material has positive $Be$ and is therefore 
gravitationally unbound.
This material forms powerful bipolar outflows.
In Models~G and J with smaller viscosity
($\alpha\la 0.1$) almost all matter that outflows through the outer boundary
remains bound ($Be<0$) and cannot escape to large radial distances.
\label{fig2}}
\end{figure}

\placefigure{fig2}

As we explained in Introduction, the numerical models of ADAFs
could be divided into four types that are characterized 
by different flow patterns.
We shall now describe the properties of the flows of the various types.

\subsubsection{Bipolar outflows}

The high viscosity ($\alpha=1$) Models~A, B, C 
and moderate viscosity ($\alpha=0.3$) Model~D
are stationary, symmetric
with respect to the equatorial plane, and show a flow pattern with
equatorial inflow and bipolar outflows. 
In Models~A, B and C 
the mass is strongly concentrated towards the equatorial plane, 
and the flow patterns depend weakly on $\gamma$.
The angular momentum is significantly smaller than
the Keplerian one, and the pressure gradient force plays
a major role in balancing gravity. 
We discuss here two representative high viscosity Models~A ($\gamma=5/3$) 
and C ($\gamma=4/3$). Model~B has properties somewhat between 
those of Models~A and C.

Figures~3 and 4 present selected properties of Models~A and C,
respectively, in the meridional
cross-section. Four panels in each figure show the distributions of density
$\rho$ (upper left), pressure $P$ (upper right), 
momentum vectors $\rho\vec{v}$ multiplied by $r$ (lower left), and 
Mach number ${\cal M}=\sqrt{v_r^2+v_\theta^2}/\sqrt{\gamma} c_s$ (lower right).
The correspondent distributions in Figures~3 and 4 are almost identical 
except the distributions of ${\cal M}$. 
In Model~A (as well as in Models~B and D) 
the flow is everywhere subsonic up to
the inner absorbing boundary at $3 r_g$.
In the equatorial inflow the radial profile of ${\cal M}$ is flat and
reaches the maximum value ${\cal M}\simeq 0.7$.
In Model~C the equatorial inflow is supersonic in a large range
of radii inside about $3\times 10^3 r_g$. The equatorial values
of ${\cal M}(r)$ increase with decreasing radius and take the maximum
value ${\cal M}=2.4$ at the inner boundary.
Thus, the presence of the supersonic or subsonic inflow 
in high viscosity models depends on the value of $\gamma$,
which controls the `hardness' of the equation of state (2.4). 
We note that in purely radial supersonic inflows 
the viscous torque cannot be 
efficient because of a reduction of the upstream transport of 
viscous interactions. This effect was referred as a `causal' effect 
(Popham \& Narayan 1992). Due to this effect the supersonic inflow in
geometrically thin and centrifugally supported accretion disks 
is possible only in the innermost region with radius close to the radius of 
the last stable black hole orbit $r_s$. However, in the case
of Model~C the viscous interaction between the supersonic equatorial
inflow and the outflowing material plays an important role.
The expansion velocities in the outflows are subsonic, and
there is no problem with causality. In this case 
the inflow can be supersonic for radii $r\gg r_s$.

\begin{figure}
\plotone{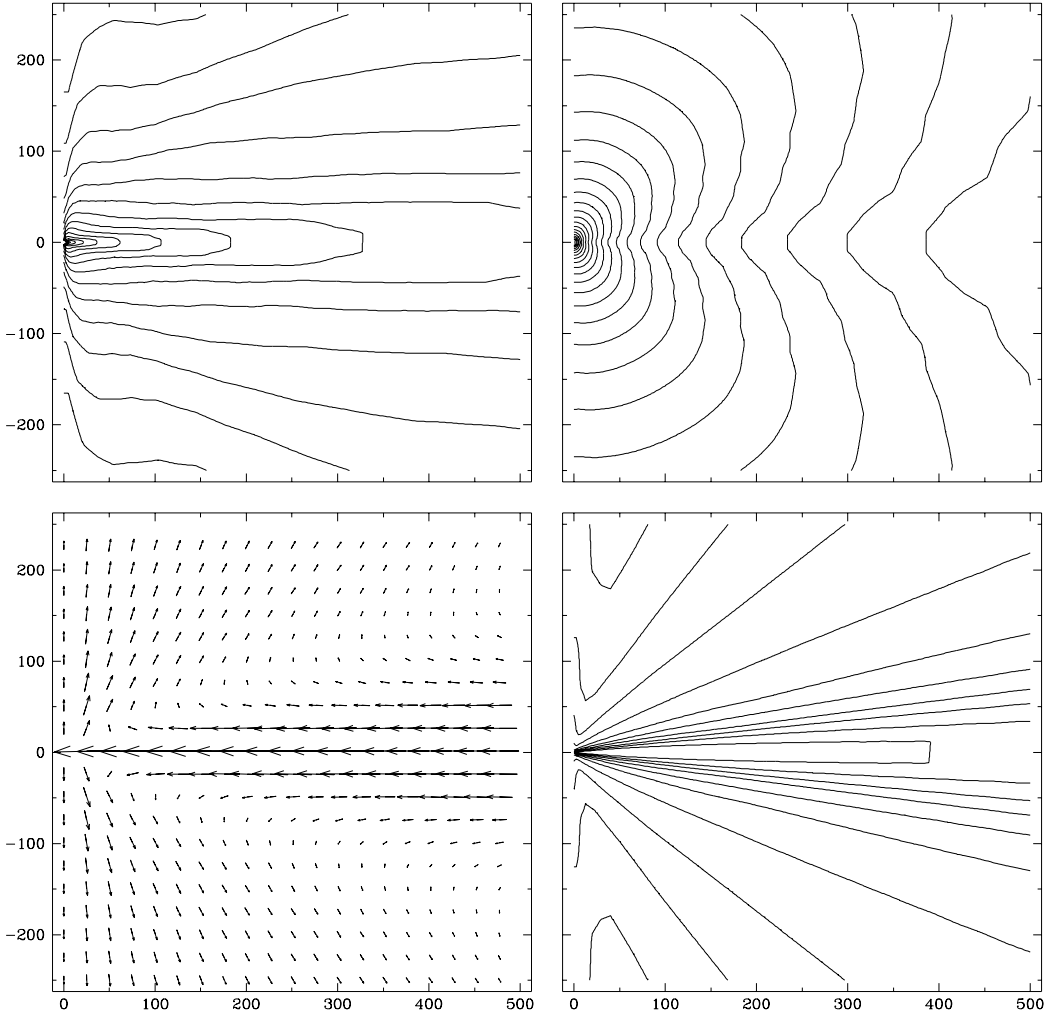}
\caption{Some selected properties of Model~A ($\alpha=1$, $\gamma=5/3$
and $Pr=\infty$) in the meridional cross-section. 
The vertical axis coincides with the axis 
of rotation. The black hole is located in the origin. 
The axes are labeled in units
of $r_g$. Only the inner region of the extended computational domain
with the outer boundary located at $r_{out}=8000 r_g$ is shown. 
The upper left panel shows the contours of density $\rho$.
The contour lines are spaced with $\Delta\log\rho=0.2$. The density
monotonically increases toward the black hole. 
Upper right: the contours of pressure $P$. The lines are spaced with
$\Delta\log P=0.2$. The pressure 
monotonically increases toward the black hole.
Lower left: the momentum vector field scaled by $r$. The direction and 
length (in relative units) of an arrow correspond to
those of the vector $r\rho\vec{v}$.
The flow pattern consists of the equatorial inflow and bipolar outflows.
Lower right: the contours of the Mach number ${\cal M}$. 
The lines are spaced with
$\Delta{\cal M}=0.1$. The maximum ${\cal M}$ at a given radius is reached
at the equatorial plane. The flow is subsonic (${\cal M}<1$) everywhere
in our computation domain with the maximum value ${\cal M}\simeq 0.7$. 
\label{fig3}}
\end{figure}

\begin{figure}
\plotone{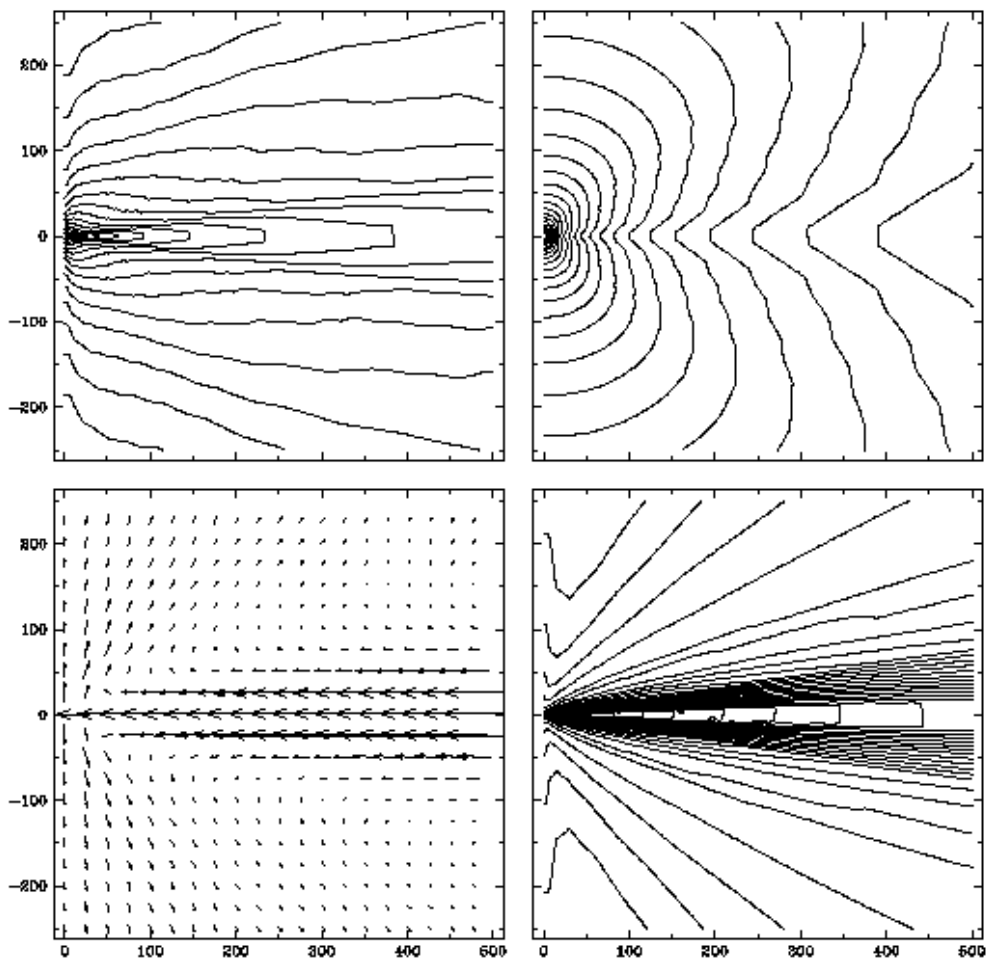}
\caption{Some selected properties of Model~C ($\alpha=1$, $\gamma=4/3$
and $Pr=\infty$) in the meridional cross-section.
The difference of this model from Model~A
presented in Figure~3 appears mostly
in the distributions of the Mach number ${\cal M}$
(lower right panels). Model~C shows supersonic 
(${\cal M}>1$) equatorial inflow.
The thick line in the distribution of ${\cal M}$ corresponds to the value
${\cal M}=1$. Other details as in Figure~3.
\label{fig4}}
\end{figure}

\placefigure{fig3}
\placefigure{fig4}

To quantitatively characterize the process of bipolar outflows we have
estimated the `mass inflow rate' $\dot{M}_{in}(r)$ by adding up all the
inflowing gas elements (with $v_r<0$) at a given radius $r$, 
and compared $\dot{M}_{in}$ with the net accretion rate $\dot{M}_{0}$.
The results are shown in Figure~5 for Models~A
(solid line), C (dotted line) and D (dashed line).
The curves do not show power-law dependences,
because they do not look like straight lines on the log--log plot.
From Figure~5 one can see that $\dot{M}_{in}$,
and correspondingly the `mass outflow rate', 
$\dot{M}_{out}=\dot{M}_{in}-\dot{M}_{0}$,
strongly depend on $\gamma$ at a fixed $\alpha$.
In the case of Model~A only about $1/7$ of matter that inflows
at $r=1000 r_g$ reaches the black hole, whereas in the case of Model~C 
the fraction is about $1/2$.
Model~D shows that a reduction of $\alpha$ by three times, 
with respect to the one for Model~A, results in
$\simeq 2-3$ times suppression of the mass outflow rate $\dot{M}_{out}$
in the radial range of $\sim 10^2-10^3 r_g$.
Models with smaller $\gamma$ show the tendency of
larger suppression of $\dot{M}_{out}$ with decreasing $\alpha$. 

\begin{figure}
\plotone{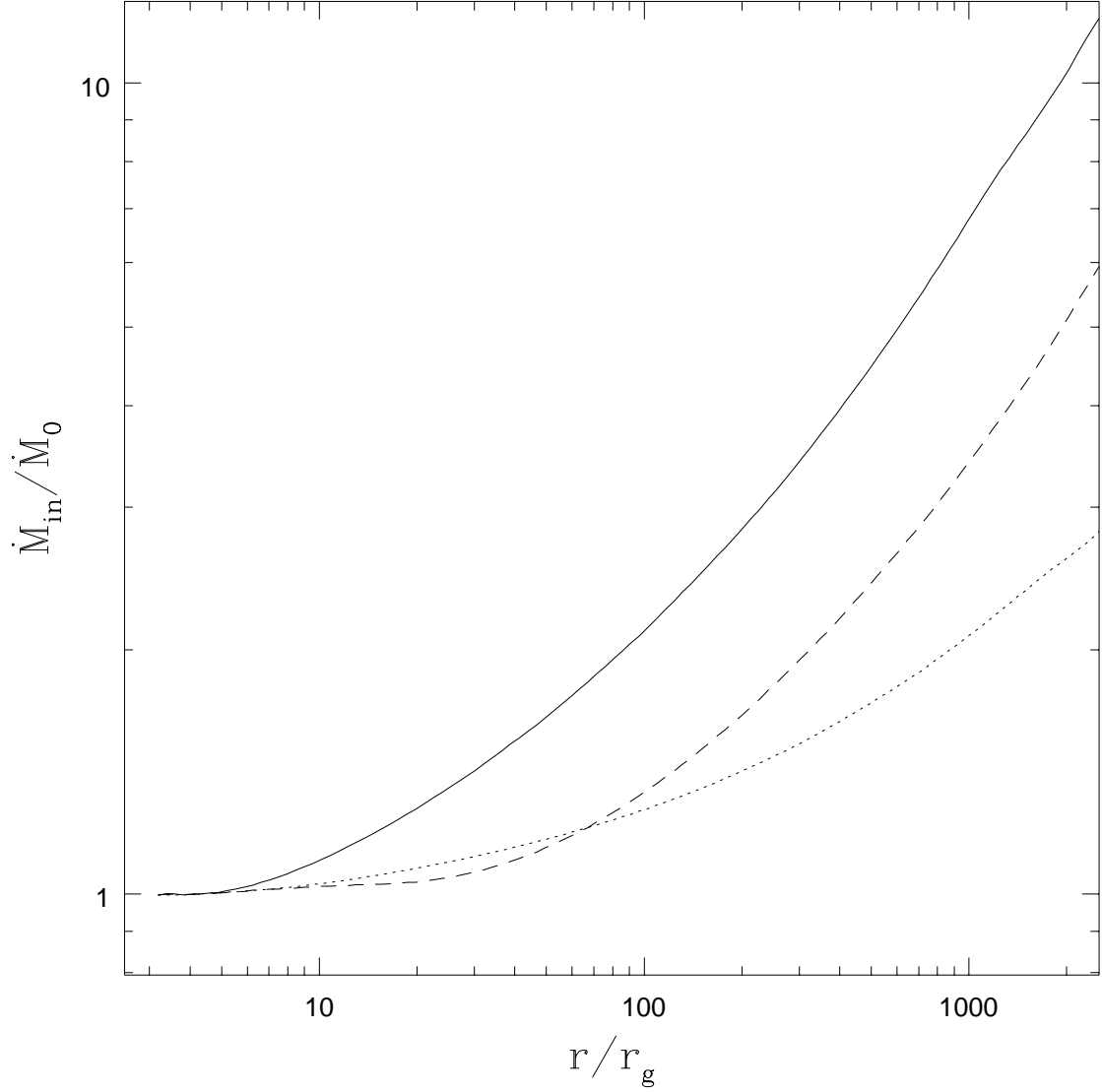}
\caption{The mass inflow rate $\dot{M}_{in}$
as a function of radius in three stationary 
models with bipolar outflows: Model~A (solid line),
Model~C (dotted line) and Model~D (dashed line).
The values of $\dot{M}_{in}$ are calculated by adding up
all the inflowing gas elements at a given radius $r$.
$\dot{M}_{in}$ is normalized to the net accretion rate $\dot{M}_{0}$.
The mass outflow rate $\dot{M}_{out}=\dot{M}_{in}-\dot{M}_0$.
\label{fig5}}
\end{figure}

\placefigure{fig5}

It is interesting to compare  the behaviour of the dimensionless quantities
$$
\lambda={1\over \dot{M}_{in}\ell_K}
\left(\dot{M}_{in}\ell+2\pi r^3 \int \Pi_{r\phi}\sin\theta d\theta\right),
\eqno(3.2)
$$
and
$$
\epsilon={1\over \dot{M}_{in}v_K^2}
\left(\dot{E}_{adv}+\dot{E}_{visc}\right), \eqno(3.3)
$$
in our models with prediction of self-similar solutions in which
$\lambda$ and $\epsilon$ are constants (see Blandford \& Begelman 1999).
In (3.2) and (3.3) we have used the following notation,
$$
\dot{E}_{adv}=2\pi r^2 \int\rho v_r\left({v^2\over 2}+W+\Phi\right)
\sin\theta d\theta, \eqno(3.4)
$$
$$
\dot{E}_{visc}=2\pi r^2 \int(v_r\Pi_{rr}+v_\theta\Pi_{r\theta}+
v_\phi\Pi_{r\phi})\sin\theta d\theta, \eqno(3.5)
$$
$\ell=v_\phi r\sin\theta$ is the
specific angular momentum, and $v_K=\Omega_K r$ and $\ell_K=\Omega_K r^2$ 
are the Keplerian angular velocity and angular momentum, respectively. 
Integration in (3.2)-(3.5) has been taken over those
angles $\theta$ for which $v_r<0$. In our models $\lambda$ and $\epsilon$
are changed with radius. The functions $\lambda(r)$ and $\epsilon(r)$ 
are plotted in Figure~6 by the solid,
dotted and dashed lines for Models~A, C and D, respectively.
Analysing the dependences in Figure~6 as well as the behaviour
of $\dot{M}_{in}$ in Figure~5, one conclude that
Models~A, C and D do not reveal the self-similar behaviour.

\begin{figure}
\plotone{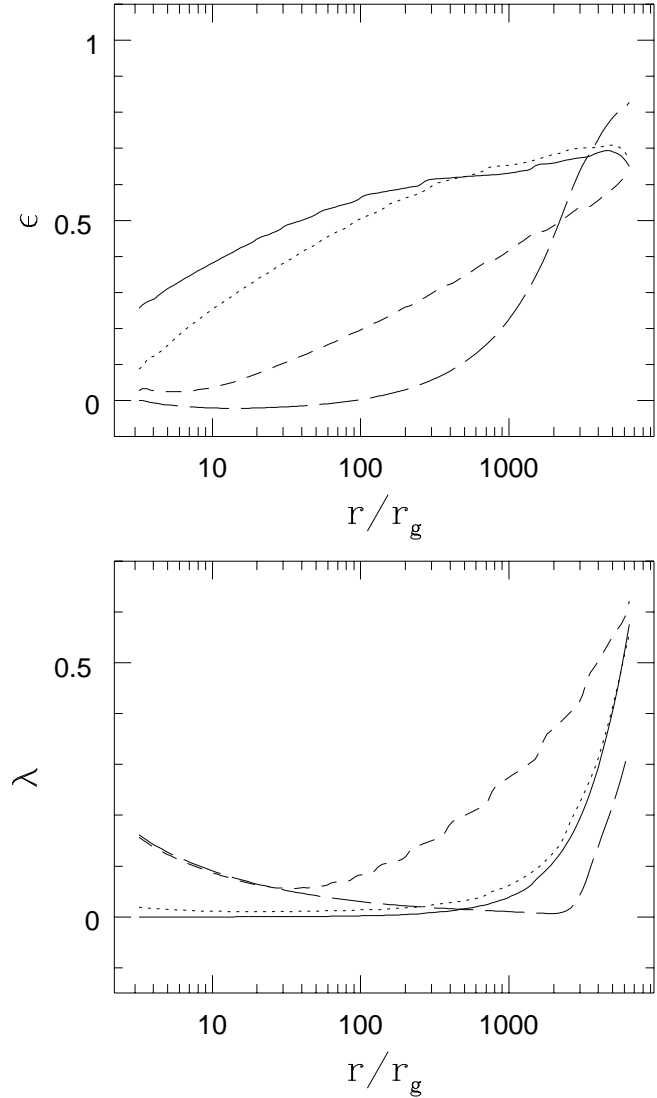}
\caption{Radial dependences of the dimensionless quantities $\epsilon$ and
$\lambda$ defined by equations (3.2), (3.3) for a variety of the moderate
and high viscosity models. 
The solid, dotted, dashed and long-dashed lines 
in both panels correspond to
Models~A, C, D and E, respectively.
Self-similar solutions predict $\epsilon$ and $\lambda$  to be constant. 
Only in pure inflow Model~E (long-dashed lines)
$\epsilon$ is approximately constant ($\epsilon\approx 0$) at $r\la 10^3 r_g$
and the model can be considered as the self-similar 
ADAF. Other models do not reveal self-similar behaviour.
\label{fig6}}
\end{figure}

\placefigure{fig6}

Figures~7 and 8 show the angular structure (in the $\theta$-direction)
in Model~A and D, respectively, at four radial positions of
$r=30 r_g$ (long-dashed lines), $100 r_g$ (dashed lines), $300 r_g$
(dotted lines) and $1000 r_g$ (solid lines).
%Plots show close qualitative coincidence of flow structures in both models.
The angular distribution of density demonstrates a 
considerable concentration of matter
towards the equatorial plane in high viscosity Model~A. 
The concentration is less significant in the case of moderate 
viscosity Model~D.
The models show quite flat distributions of angular velocity
$\Omega$, especially at smaller $r$. The radial velocities $v_r$ are negative
in the equatorial inflowing regions, 
where the mass concentration takes place.
The wide polar regions are filled by the unbound
outflowing matter with positive $v_r\ga v_K$.
The polar outflows are less effectively accelerated
in Model~D, and it results in a reduction of the
mass outflow rates $\dot{M}_{out}$ in comparison with those
of Model~A (dashed and solid lines in Figure~5, respectively).
%Outflow velocities are sufficiently higher in high viscosity Model~A.
In the polar regions $c_s/v_K\ga 1$, whereas in the inflowing part 
one has $c_s/v_K<1$.
Note, that the ratio $c_s/v_K$ equals to the
relative thickness of the accretion flow in the vertically averaged
accretion disk theory (e.g. Shakura \& Sunyaev 1973), $h/r=c_s/v_K$. 
In accretion disks one has $h/r\la 1$. The case $h/r\ga 1$ corresponds to 
a thermally expanded unbound gas cloud.
Models~B and C demonstrate angular structures qualitatively
similar to that of Model~A.

\begin{figure}
\plotone{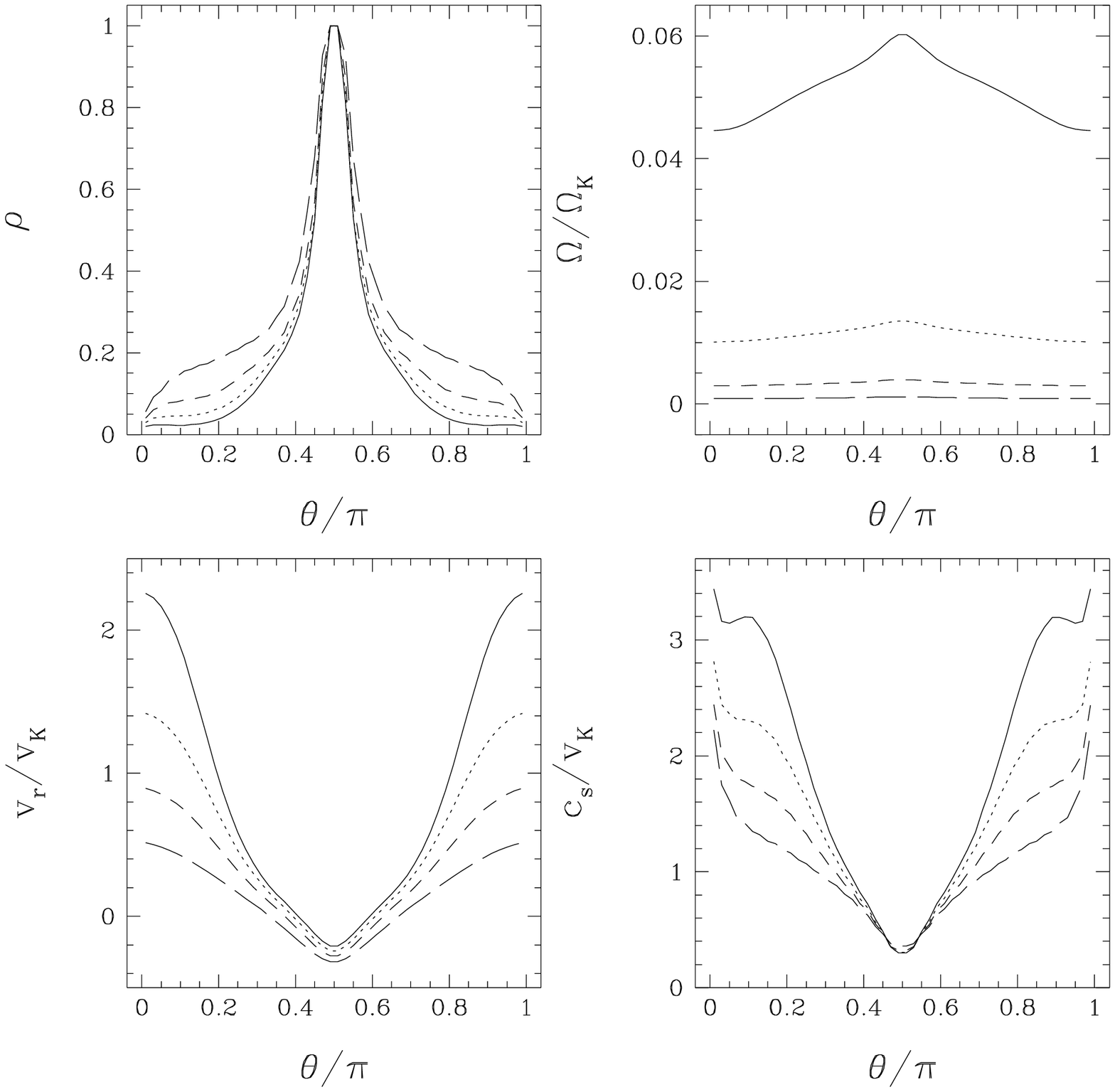}
\caption{Angular profiles of the density $\rho$, angular velocity $\Omega$,
radial velocity $v_r$ and sonic velocity $c_s$
from Model~A  at four radial positions of
$r=30 r_g$ (long-dashed lines), $100 r_g$ (dashed lines), $300 r_g$
(dotted lines) and $1000 r_g$ (solid lines). The values of $\rho$ have been
normalized to the maximum value of $\rho$ at the corresponding radius.
\label{fig7}}
\end{figure}

\begin{figure}
\plotone{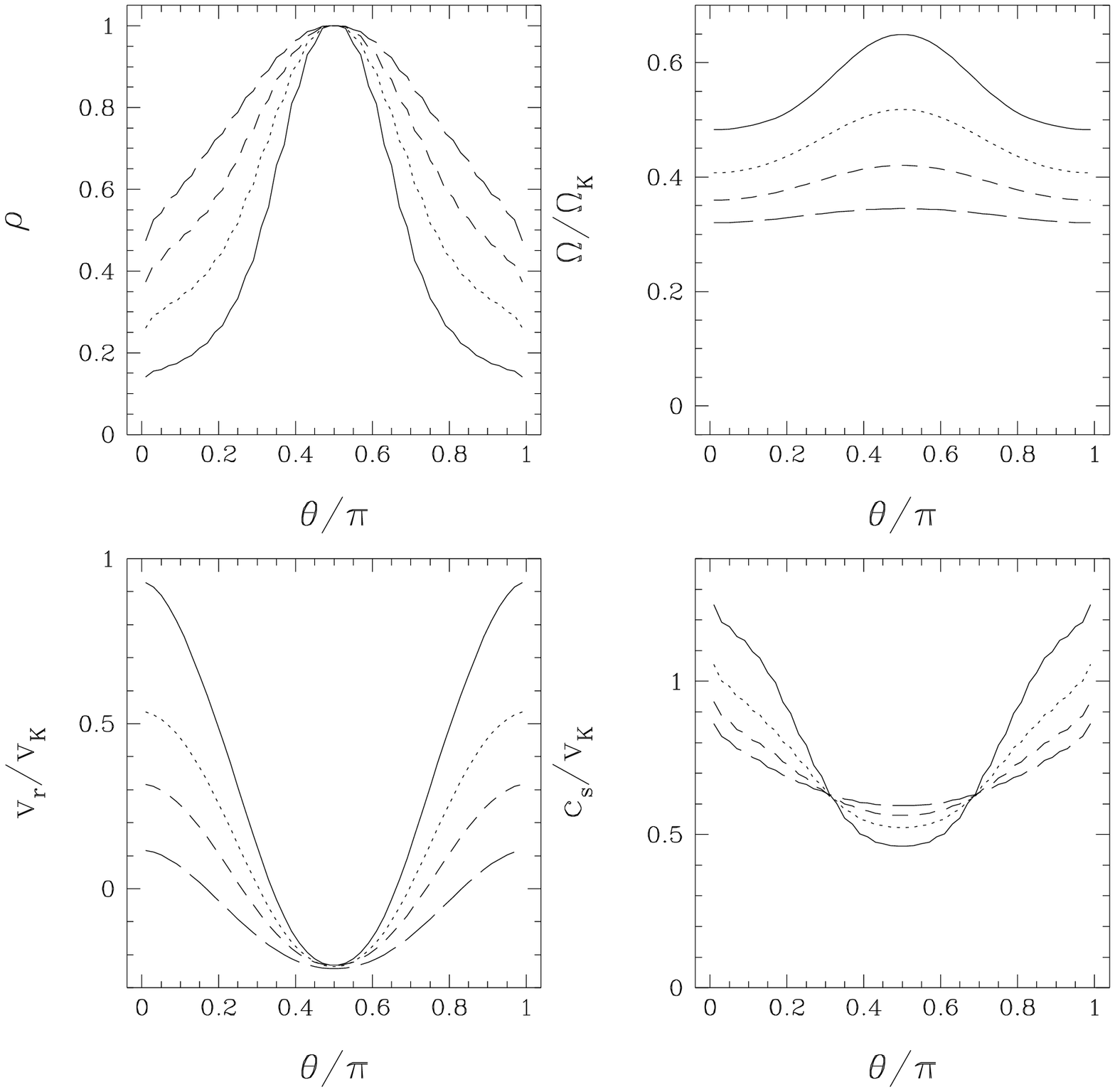}
\caption{Angular profiles of the density $\rho$, angular velocity $\Omega$,
radial velocity $v_r$ and sonic velocity $c_s$
from Model~D  at four radial positions of
$r=30 r_g$ (long-dashed lines), $100 r_g$ (dashed lines), $300 r_g$
(dotted lines) and $1000 r_g$ (solid lines). The values of $\rho$ have been
normalized to the maximum value of $\rho$ at the corresponding radius.
\label{fig8}}
\end{figure}

\placefigure{fig7}
\placefigure{fig8}

Figure~9 shows the averaged radial structure for a variety of models. 
Models~A, C and D are represented by the solid, dotted 
and dashed lines, respectively.
All plotted quantities, $\rho$, $\Omega$, $v_r$, $c_s$,
have been averaged over the polar angle
$\theta$ with the weighted function $\rho$, except for $\rho$ itself.
The density and sonic velocity profiles
in the models can approximately be described by 
radial power-laws, with $\rho\propto r^{-1}$
and $c_s\propto r^{-1/2}$. 
The radial velocity is about $v_r\propto r^{-1}$ 
in the case of Models~A and D,
but does not show any distinguished power-law dependence 
in the case of Model~C.
The latter model shows faster increase of the radial velocity with
decreasing radius.
The most significant difference between the high and moderate viscosity 
models can be seen in the radial profiles of $\Omega$.
The angular velocity $\Omega$ shows quite unexpected behaviour in 
the case of high viscosity Models~A and C.
In Model~A, $\Omega\propto r^{-1/2}$ in all ranges of the radii.
This radial dependence is significantly flatter than the one for
the Keplerian angular velocity, $\Omega_K\propto r^{-3/2}$.
In Model~C, $\Omega\propto r^{-1/2}$ in the outer region,
at $r\ga 10^3 r_g$, and
$\Omega\propto\Omega_K\propto r^{-3/2}$ in the inner region,
at $r\la 10^2 r_g$.
The steeper dependence of $\Omega(r)$ in the inner region of Model~C could be
due to the equatorial supersonic inflow (see Figure~4, lower right panel).
In moderate viscosity Model~D, the angular velocity profile is not surprising, 
it is quite close to the Keplerian one.

\begin{figure}
\plotone{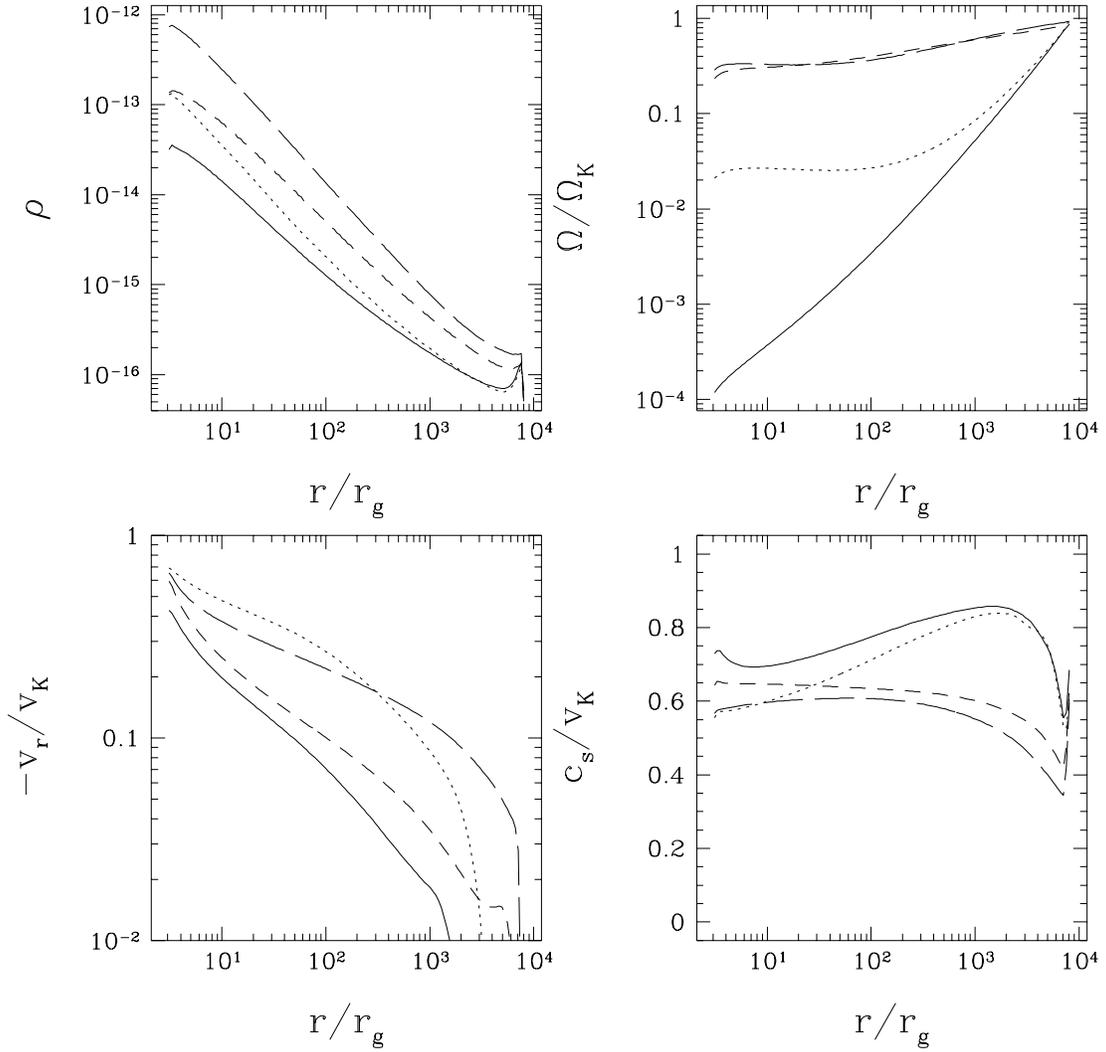}
\caption{Radial structure of the flow in Models~A (solid lines),
C (dotted lines), D (dashed lines) and E (long-dashed lines).
All plotted quantities -- the density $\rho$,
angular velocity $\Omega$, radial velocity $v_r$
and sonic velocity $c_s$ -- have been averaged over the polar angle $\theta$.
\label{fig9}}
\end{figure}

\placefigure{fig9}

\subsubsection{Pure inflows}

As we noted earlier, Model~E demonstrates some peculiar properties.
The model is stable and does not form outflows
except very close to the outer boundary, at $r\ga 4000 r_g$. 
Figure~10 shows some selected properties
of Model~E in the meridional cross-section.
The flow pattern looks very similar to the one 
for the spherical accretion flow.
However, contrary to spherical flows, the rotating accretion flow 
in Model~E has a reduced inflow rate at the equatorial and polar regions 
(compare vectors in the lower left panel of Figure~10) 
and a corresponding local decrease of Mach number
there (Figure~10, lower right panel).
The net radial energy flux is close to zero for $r\la 10^3 r_g$
in this model; the inward advection of energy balances the outward-directed 
energy flux due to viscosity (see plot for $\epsilon$ in Figure~6, 
long-dashed line).
Such a balance of the inward and outward energy fluxes in Model~E
coincides with a property
of the self-similar ADAF solutions, in which
$\epsilon=0$. An other property of the self-similar ADAFs is that $\lambda=0$.
The latter property is a consequence of the assumption that
the inner boundary located at $r=0$, and there is a zero
outward flux of angular momentum. In reality, however,
the inner boundary locates at a finite radius, and there must be a non-zero 
outward flux of angular momentum $\approx\dot{M}_{0}\ell(r_{in})$.
In this case $\lambda$ is a function of radius,
$\lambda(r)\propto\ell_K^{-1}\propto r^{-1/2}$.
Model~E confirms the latter dependence in a wide range of radii,
at $r\la 2\times 10^3 r_g$, as can be seen from
the $\lambda(r)$ plot in the lower panel of Figure~6 (long-dashed line).
It is interesting to note that Model~D demonstrates
a similar behaviour of $\lambda(r)$
(dashed line in the lower panel of Figure~6)
in the innermost part, at $r\la 30 r_g$, where
the outflow rate $\dot{M}_{out}$ is small (see Figure~5).

\begin{figure}
\plotone{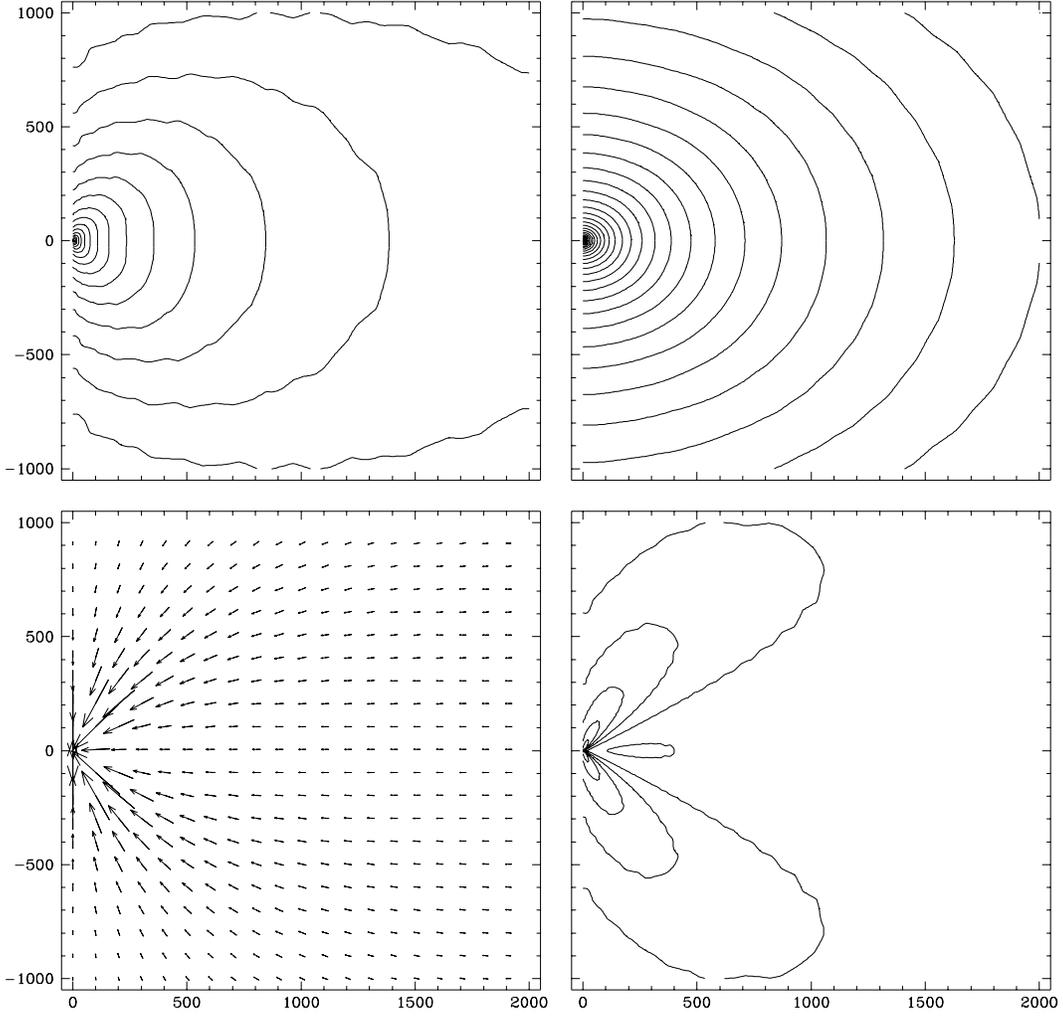}
\caption{Some selected properties of Model~E ($\alpha=0.3$, $\gamma=3/2$
and $Pr=\infty$) in the meridional cross-section.
See the caption of Figure~3 for details.
Distributions of density (upper left) and pressure (upper right)
are quite close to the spherical one. The mass inflow rate in the
equatorial region smaller than the one in the polar regions (lower left).
The decrease of Mach number at the
equatorial region (lower right) is connected with the reduction
of the inflow rate.
\label{fig10}}
\end{figure}

\placefigure{fig10}

The radial and angular structure of Model~E can be seen in Figures~9
(long-dashed lines) and 11. 
The radial profiles of the $\theta$-averaged $\rho$, $\Omega$ and $c_s$ 
are very close to those for the self-similar ADAFs:
$\rho\propto r^{-3/2}$, $\Omega\propto r^{-3/2}$ and $c_s\propto r^{-1/2}$.
But, the angular profiles of them do not correspond to any of 
the two-dimensional self-similar solutions
found by Narayan \& Yi (1995a). The discrepancy is connected with
a reduction of mass inflow rate in the equatorial region 
clearly seen in Model~E.
The parameter $Be$ is positive in the inner region of Model~E, 
at $r\la 10^3 r_g$, and there are no outflows. 

\begin{figure}
\plotone{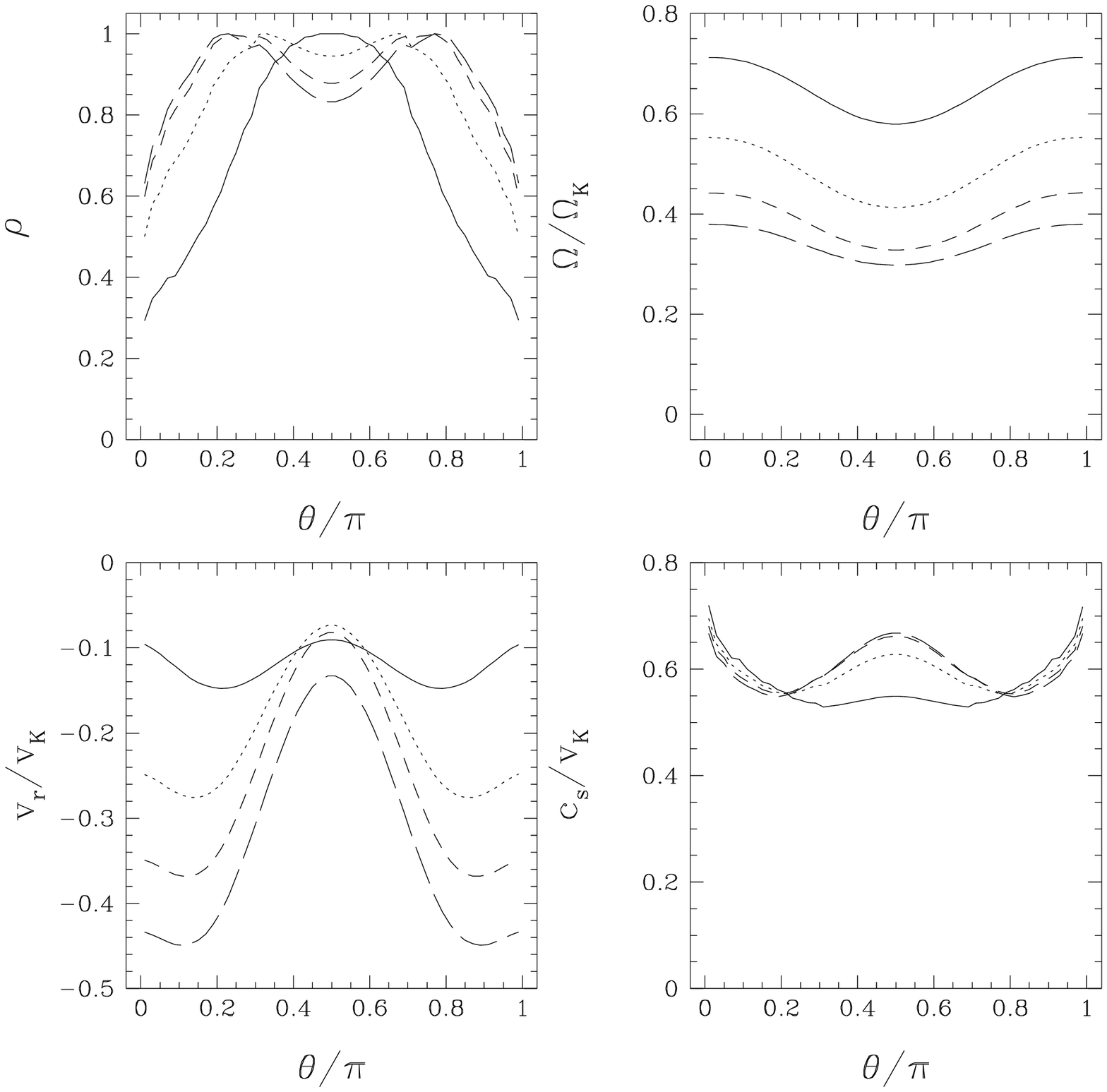}
\caption{Angular profiles of the density $\rho$, angular velocity $\Omega$,
radial velocity $v_r$ and sonic velocity $c_s$
from Model~E  at four radial positions of
$r=30 r_g$ (long-dashed lines), $100 r_g$ (dashed lines), $300 r_g$
(dotted lines) and $1000 r_g$ (solid lines). The values of $\rho$ have been
normalized to the maximum value of $\rho$ at the corresponding radius.
\label{fig11}}
\end{figure}

\placefigure{fig11}

\subsubsection{Large-scale circulations}

Models~F, G, H and I with large-scale circulations have
a moderate viscosity, $0.1\la\alpha\la 0.3$.
In the ($\alpha$, $\gamma$) plane they locate
on both sides of the line that separates the stable and unstable flows
(see Figure~1, left panel).
Below we discuss representative stable Model~G and 
unstable Models~F and I.

The flow pattern for stable Model~G
is presented in Figure~12.
In the lower left panel of Figure~12 
one can see the inner part of a meridional cross-section of the global
circulation cell which has a torus-like form in three dimensions.
The polar outflow in the upper hemisphere becomes supersonic from
$\sim 1000 r_g$ outward. The polar funnel in the lower hemisphere
filled up by low-density matter is clearly seen in the distributions of
density (upper left), pressure (upper right) and Mach number (lower right).
The low-density matter in the funnel forms an accretion flow at small $r$ and
an outflow at larger $r$. The boundary between these inflowing and outflowing
parts is variable and determines 
the supersonic/subsonic accretion regime in the funnel. 
At the particular moment shown in Figure~12 
the boundary between the inflow and outflow
in the funnel is located at small $r$ and 
the accretion flow is mostly subsonic.
Figure~13 shows the angular structure of the flow in Model~G. 
The equatorial asymmetry of the flow pattern 
introduces the asymmetry in the angular profiles of all quantities shown.
Note the impressive similarity of the profiles at different radial distances
except regions close to the poles.

\begin{figure}
\plotone{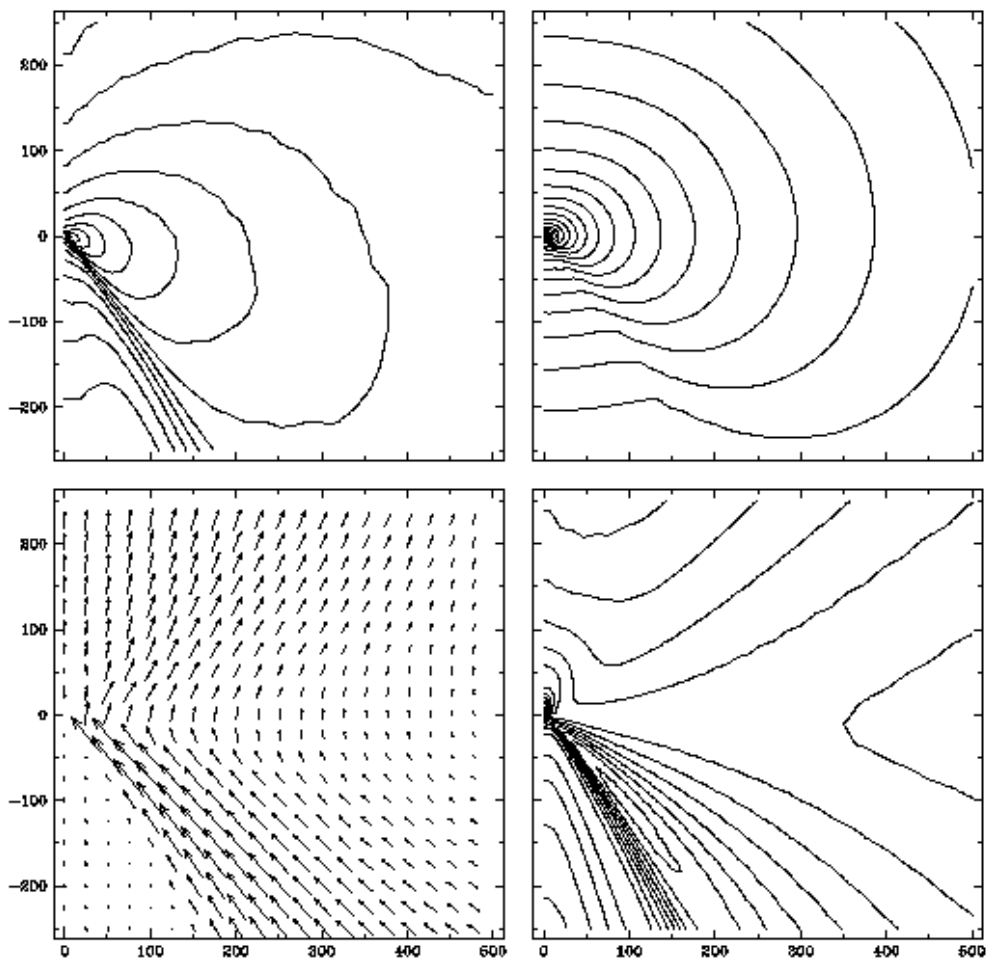}
\caption{Some selected properties of Model~G ($\alpha=0.1$, $\gamma=5/3$
and $Pr=\infty$) in the meridional cross-section.
See the caption of Figure~3 for details.
The inner part of the global circulation cell with a unipolar outflow directed
to the upper hemisphere is seen in momentum vectors (lower left).
The polar funnel in the lower hemisphere is filled by low density matter.
The thick line in the distribution of 
the Mach number (lower right) corresponds to
the maximum contour ${\cal M}=0.8$.
\label{fig12}}
\end{figure}

\begin{figure}
\plotone{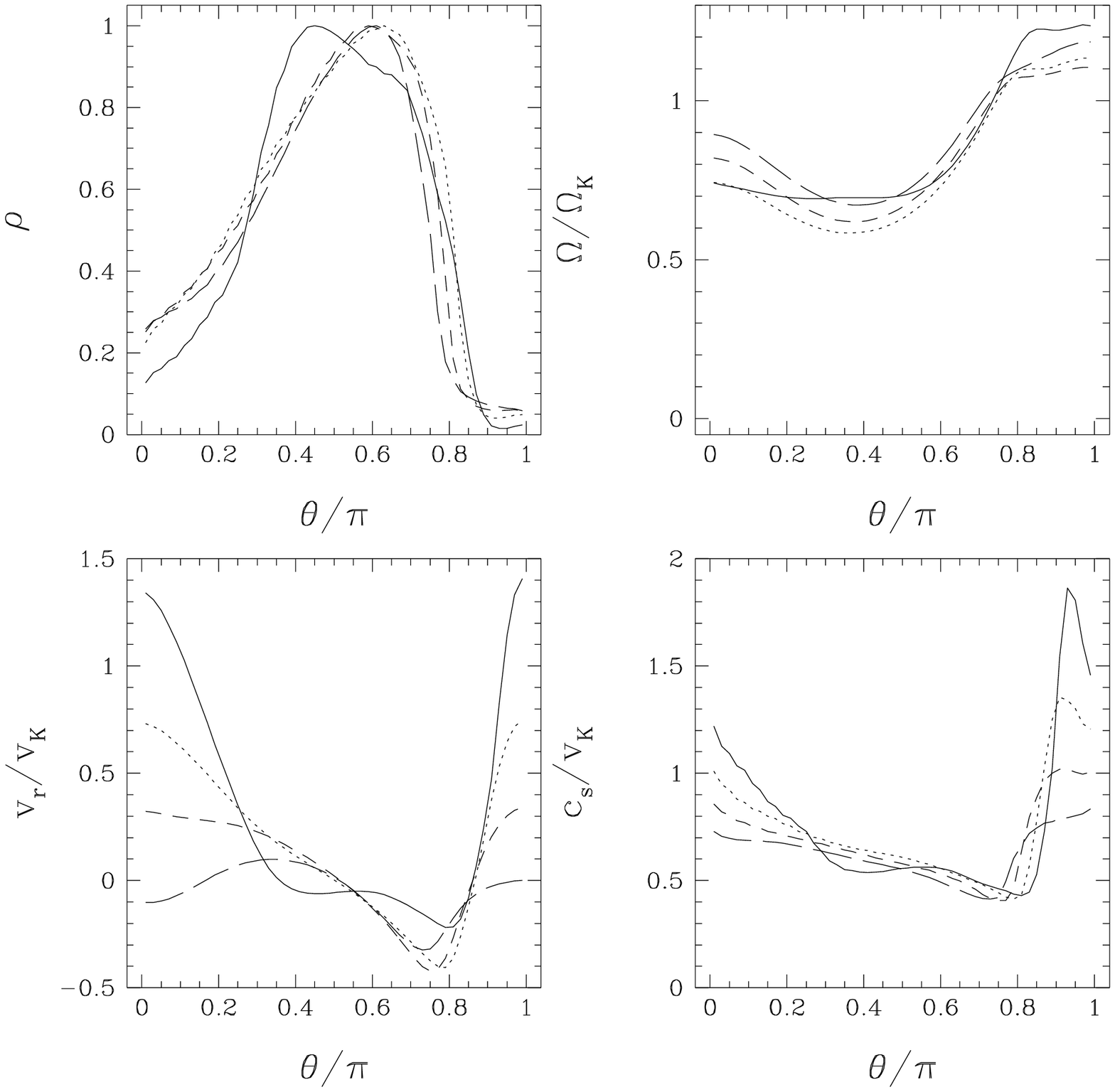}
\caption{Angular profiles of the density $\rho$, angular velocity $\Omega$,
radial velocity $v_r$ and sonic velocity $c_s$
from Model~G  at four radial positions of
$r=30 r_g$ (long-dashed lines), $100 r_g$ (dashed lines), $300 r_g$
(dotted lines) and $1000 r_g$ (solid lines). The values of $\rho$ have been
normalized to the maximum value of $\rho$ at the corresponding radius.
\label{fig13}}
\end{figure}

\placefigure{fig12}
\placefigure{fig13}

Snapshots of the unstable Models~F and I are given
in Figure~14, left and right panels, respectively. 
Model~F has a stable global flow pattern
dominated by an unipolar circulation motion.
The flow 
is quasi-periodically perturbed by growing hot convective bubbles. 
These bubbles originate in the innermost part of the accretion flow
close to the equatorial plane.
They are hotter and lighter than the surrounding matter,
and the Archimedes buoyancy forces them to move outward. 
During the motion, the bubbles are heated up further
due to viscous dissipation and migrate from the equatorial region to
the upper hemisphere. In Figure~14 (left) one can see the growing
bubble in the density contours inside $r\simeq 500 r_g$. The structure
seen in the upper polar region at $r\simeq 2000 r_g$ is a `tail' 
of the previous bubble.
Less viscous Model~I does not show a regular flow pattern.
The hot convective bubbles outflow quasi-periodically in both
the upper and lower hemispheres without a preferable direction.
Figure~14 (right) shows sequences of the convective bubbles 
in the polar regions of both hemispheres. 
The regular flow pattern with the symmetric
bipolar outflows and equatorial inflow is seen only through the time averaging
in Model~I.

\begin{figure}
\plottwo{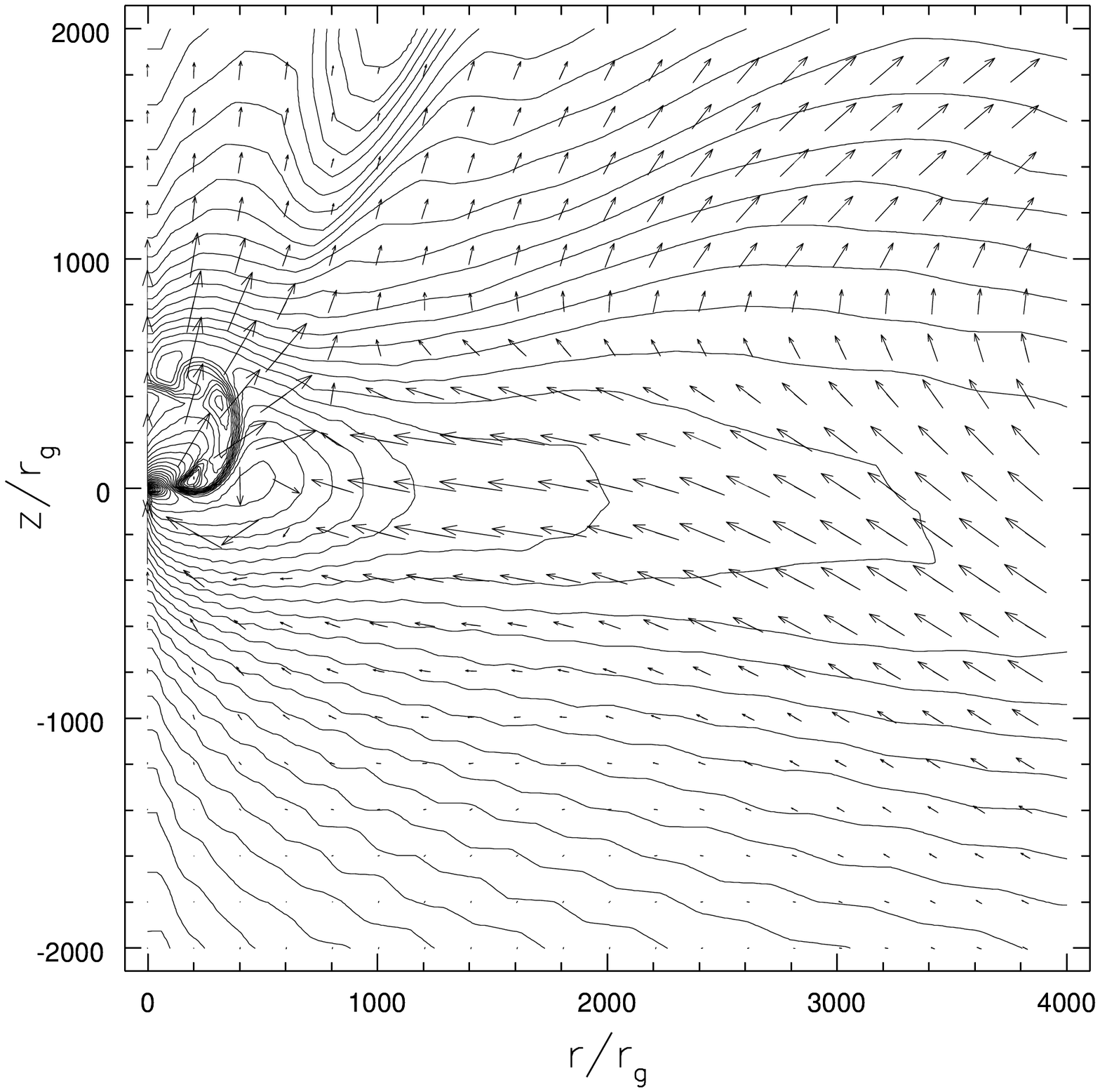}{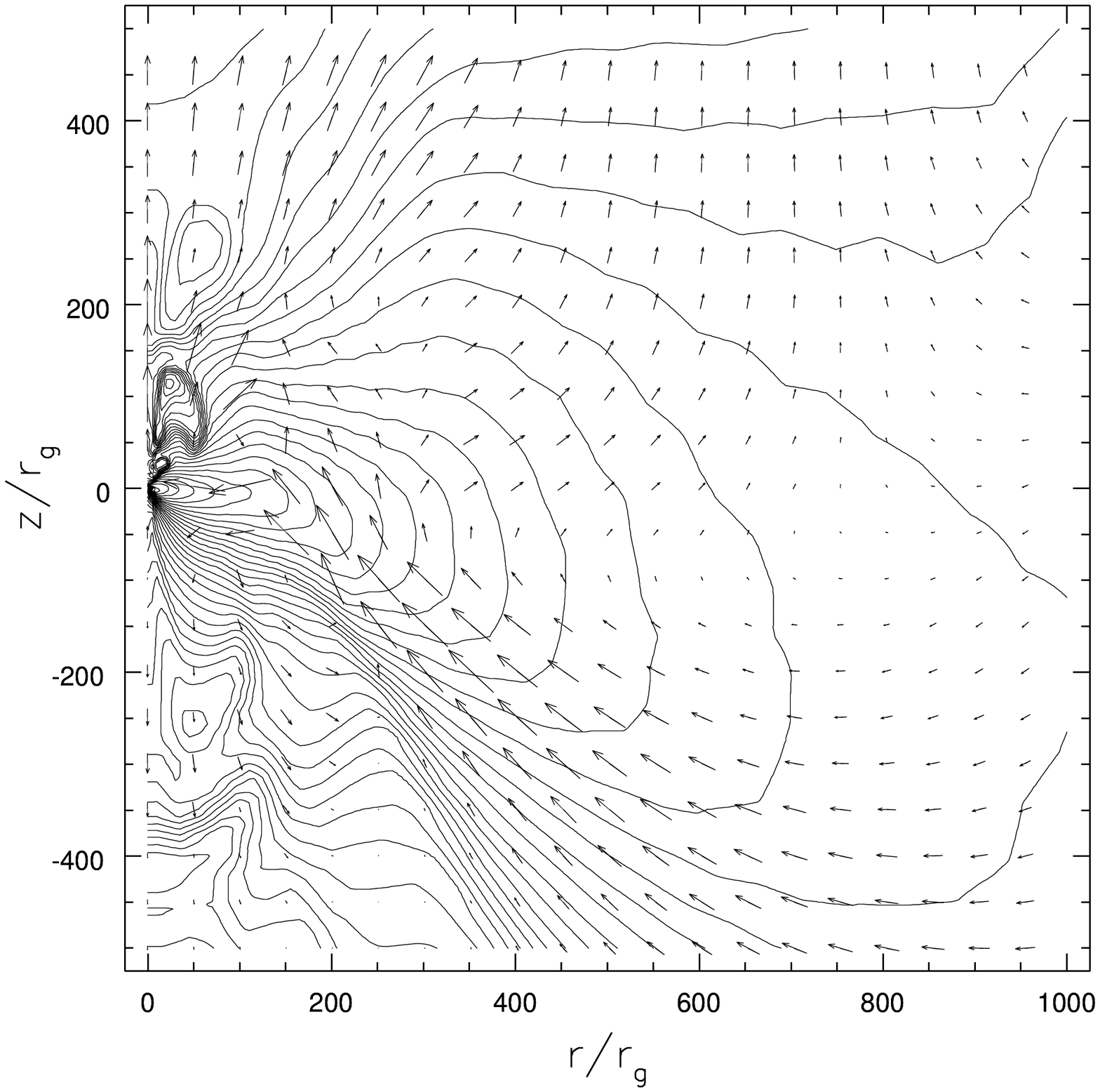}
\caption{Two snapshots of the flow structure
from unstable Models~F (left) and I (right). 
The density contours
with the step $\Delta\log\rho=0.1$ and 
the vectors $r\rho\vec{v}$ are presented.
Model~F has global circulation motions which are
quasi-periodically perturbed by growth of hot convective bubbles.
The bubbles always outflow in the upper hemisphere in Model~F.
The growing bubble is clearly seen in density contrasts inside
$r\approx 500 r_g$. The structure in the upper polar region at
$r\sim 2000 r_g$ is a `tail' of the previous bubble (left).
In less viscous Model~I the hot convective bubbles
quasi-periodically arise in the innermost region and
outflow in both polar directions. The time-averaged flow pattern in
Model~I is equatorially symmetrical.
\label{fig14}}
\end{figure}

\placefigure{fig14}

\subsubsection{Convective flows}

All our low viscosity models with $\alpha\la 0.03$ are convectively
unstable independently of $\gamma$. They show complicated 
time-dependent flow patterns
which consist of numerous vortices and circulations.
The snapshot of such a behaviour of accretion flow 
in the case of Model~M is presented in Figure~15 (left).
It demonstrates non-monotonic distributions of density and velocity in the
innermost part of the flow. 
In spite of the significant time-variability of the accretion flow
we have found a tendency towards formation of temporal coherent structures
which look like  convective cells extended in the radial direction. 
In these structures, the inflowing streams of matter are sandwiched in 
the $\theta$-direction by the outflowing streams.
These structures can be seen in the velocity vectors and in the characteristic 
radial features 
of the density distribution in the left panel of Figure~15.
The time-averaged flow patterns are smooth and do not demonstrate
small-scale features.
Figure~15 (right) shows the time-averaged distributions of the density
and momentum vectors of Model~M. In the picture
one can clearly see that the accretion is suppressed in the equatorial
sector and the mass inflows concentrate mainly along the upper and lower
surfaces of the torus-like accretion disk.

\begin{figure}
\plottwo{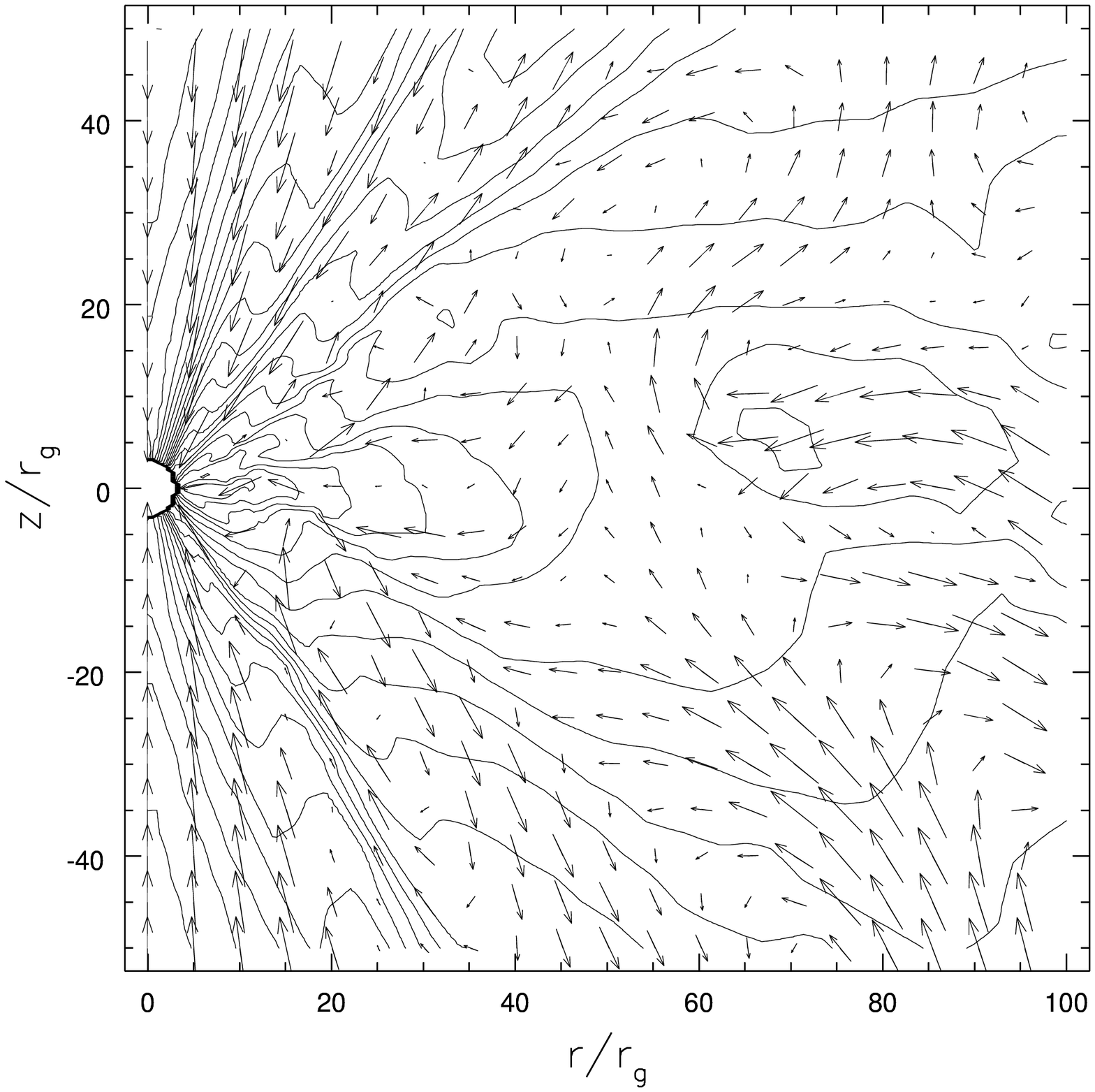}{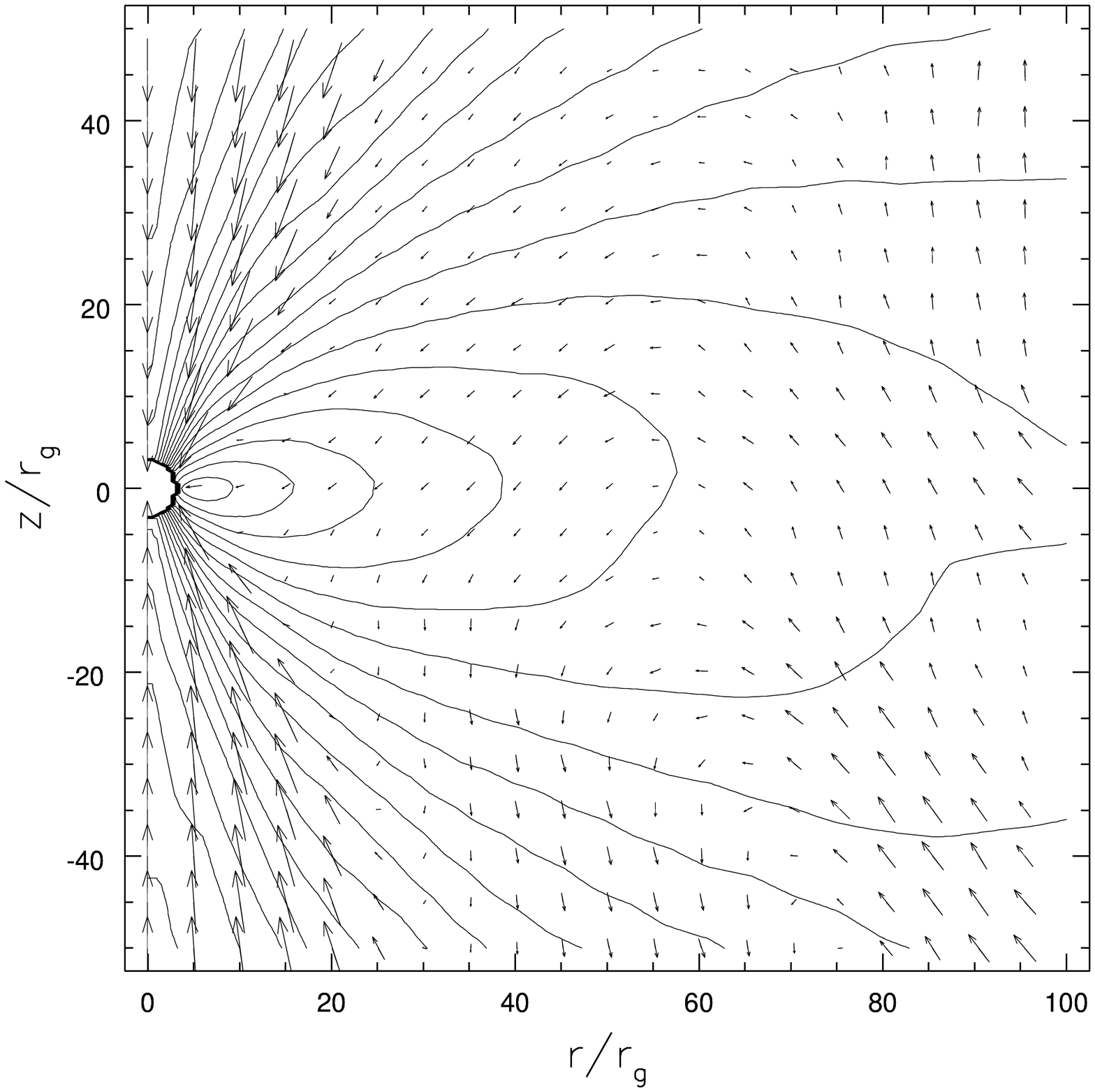}
\caption{Two-dimensional structure of the flow 
in the inner region of Model~M. The snapshot on the left
shows a non-monotonic behaviour of density (contours) and momentum flux
scaled by $r$
(arrows). The contour lines are spaced with $\Delta\log\rho=0.1$.
In the time-averaged flow pattern (right) one can clearly see
that accretion takes place mainly in the polar regions. 
The equatorial mass inflow is suppressed.
\label{fig15}}
\end{figure}

\placefigure{fig15}

Figure~16 shows the angular structure of the time-averaged flow from Model~M.
All quantities have been averaged over about $44$ periods of 
the Keplerian rotation at $r=100 r_g$.
The angular profiles of density reach their maximum values at the equator
and decrease towards the poles.
The profiles have an almost identical form
at each radius. The similarity of the angular profiles 
at different $r$ can be see as well
in the variables $\Omega$, $v_r$ and $c_s$, except in the 
regions which are close to the poles. The averaged radial velocity 
is almost zero over most of the $\theta$-range. 
It is only close to the poles that we have non-zero 
velocities (for $r\la 300 r_g$ they are negative and for larger $r$
they are positive).
The polar inflows are highly supersonic in the low viscosity models 
as can be seen by comparing the values of $v_r$ and $c_s$ from
the lower left and lower right panels of Figure~16.

\begin{figure}
\plotone{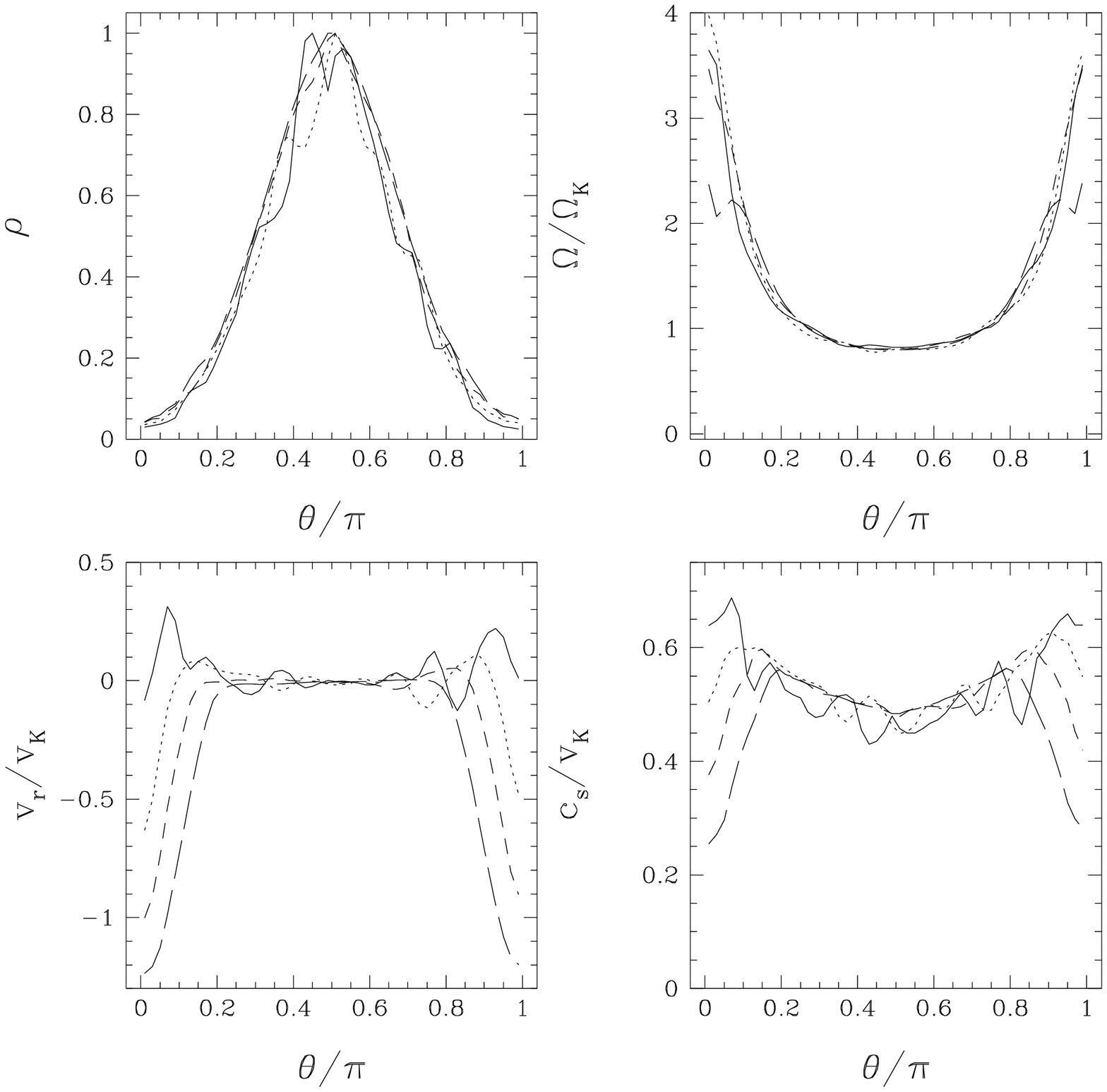}
\caption{Angular profiles of the density $\rho$, angular velocity $\Omega$,
radial velocity $v_r$ and sonic velocity $c_s$
from Model~M  at four radial positions of
$r=30 r_g$ (long-dashed lines), $100 r_g$ (dashed lines), $300 r_g$
(dotted lines) and $1000 r_g$ (solid lines). The values of $\rho$ have been
normalized to the maximum value of $\rho$ at the corresponding radius.
\label{fig16}}
\end{figure}

\placefigure{fig16}

Figure~17 shows the radial structure of the time-averaged flows in Models~J
(dotted lines), K (dashed lines), L (long-dashed lines) and
M (solid lines). It uses the same $\theta$-averaging as in Figure~9.
In all models, 
the variables $\Omega$ and $c_s$
can be described by a radial power-law, with
$\Omega\propto r^{-3/2}$ and $c_s\propto r^{-1/2}$, in the radial range
$\sim 10-10^3 r_g$.
The radial density profile can also be approximated by a power-law,
$\rho\propto r^{-\beta}$, where the index $\beta$ varies from
$\beta\approx 0.5$ for Models~J and M to $\beta\approx 0.7$ for
Models~K and L.
The radial velocities are connected to the density profiles by the relation
$v_r\propto r^{-2}\rho^{-1}\propto r^{\beta-2}$ with good accuracy.
Such a fast increase of $v_r$  inward, with respect to the free-fall velocity,
$v_{ff}\propto r^{-1/2}$, means that $v_r$ is very small at large radii
in comparison with the predictions of the `standard' 
self-similar ADAF solutions, 
for which $v_r\propto v_{ff}\propto r^{-1/2}$.
The angular velocities are close to the Keplerian ones everywhere,
and the $\Omega(r)$ profiles for Models~K, M, L have a super-Keplerian part
at the innermost region, which is typical for thick accretion disks.

\begin{figure}
\plotone{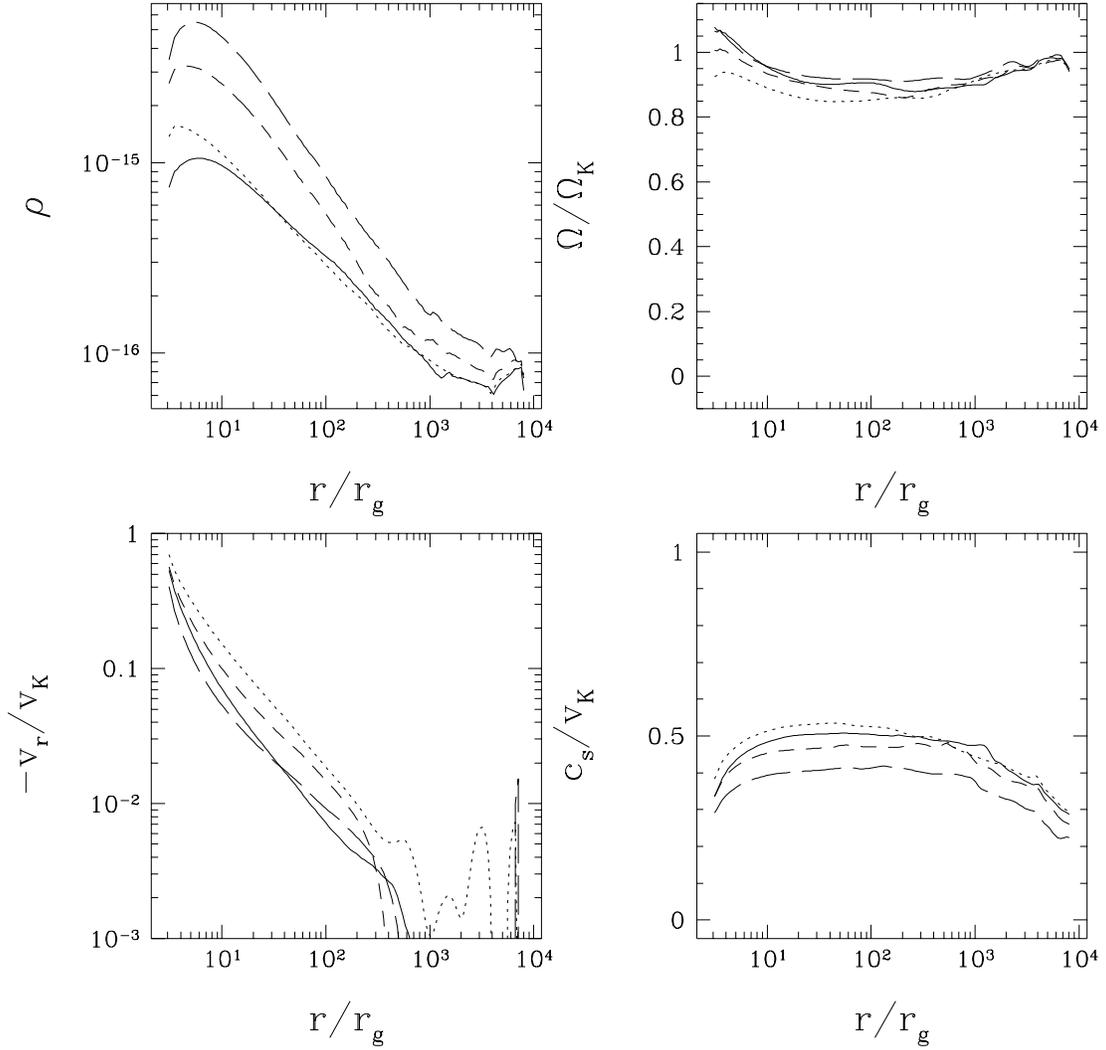}
\caption{Radial structure of the flow in Models~J (dotted lines),
K (dashed lines), L (long-dashed lines) and M (solid lines).
All plotted quantities -- the density $\rho$,
angular velocity $\Omega$, radial velocity $v_r$
and sonic velocity $c_s$ -- have been averaged over the polar angle $\theta$.
\label{fig17}}
\end{figure}

\placefigure{fig17}

The most interesting and important property of the low viscosity
models is that the convection 
transports angular momentum inward rather than outward, as it does
in the case of ordinary viscosity. The direction of 
the angular momentum transport is determined by the sign of
the ($r\phi$)-component of the Reynolds stress tensor, 
$\tau_{r\phi}=\langle v_r' v_\phi'\rangle$,
where $v_r'$, $v_\phi'$ are the velocity fluctuations and
$\langle ...\rangle$ means time-averaging. 
Negative/positive sign of $\tau_{r\phi}$
corresponds to inward/outward angular momentum transport.
Figure~18 shows the distribution of $\tau_{r\phi}$ in the meridional
cross-section of Model~M. It is clearly seen that $\tau_{r\phi}$ is
negative in most of the flow, and thus convective motions
transport angular momentum inward on the whole.

\begin{figure}
\plotone{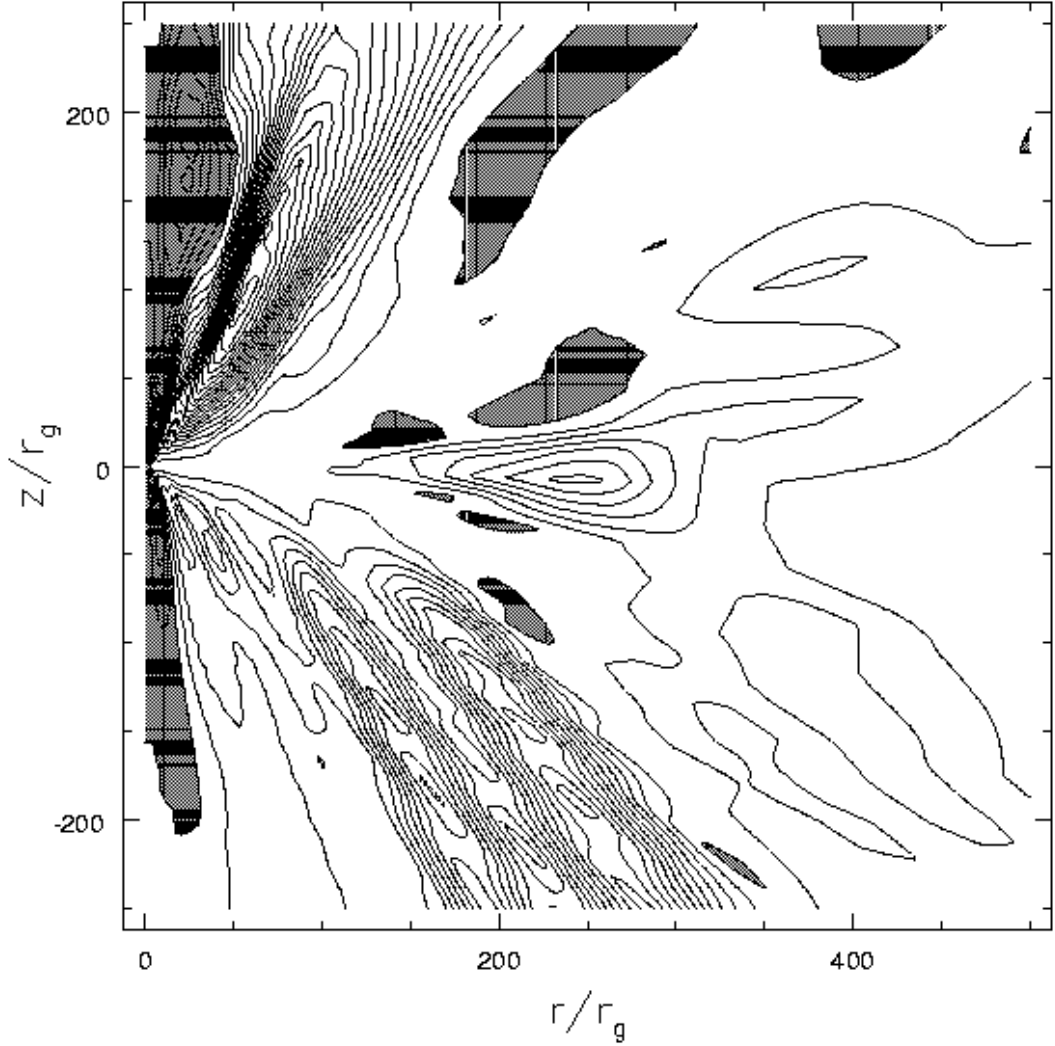}
\caption{The ($r\phi$)-component of the Reynolds stress tensor
normalized to the square of the Keplerian velocity, 
$\langle v_r' v_\phi'\rangle/v_K^2$,
in the meridional cross-section of the convectively unstable Model~K
($\alpha=0.01$, $\gamma=5/3$, $Pr=\infty$). The contours are given
with the step $\Delta=2.5\times 10^{-4}$. Regions with positive
values of the Reynolds stress are shown in grey. It is clearly seen
that the Reynolds stress is mostly negative in the bulk of the flow.
This means that the angular momentum is transported inward
by the convection motions.
\label{fig18}}
\end{figure}

\placefigure{fig18}

All low viscosity models have negative, volume averaged, $Be$
in all ranges of the radii.
Only temporal convective blobs and narrow (in the $\theta$-direction)
regions with outflowing matter
at the disk surfaces show positive $Be$.
%The  volume averaged $Be$ is negative in all range of radii.

Our numerical results for the low viscosity models 
agree with those calculated in our earlier papers (Igumenshchev et al. 1996;
Igumenshchev \& Abramowicz 1999), and are very similar to
those obtained later by Stone et al. (1999). The models 
of Stone et al. (1999) are convectively
unstable and the time-averaged radial inward velocity
is significantly reduced in the bulk of the accretion flows.
Stone et al. (1999) have checked several radial scaling laws for the viscosity
and have found the same radial dependence of the time-averaged
quantities ($\rho$, $\Omega$ and $c_s$), as those presented here, 
in the case of $\nu\propto r^{1/2}$,
which is analogous to the $\alpha$-prescription (2.5).

\subsection{Models with thermal conduction}

We have assumed the Prandtl number $Pr=1$ in the models with
thermal conduction (see Table~1).
The thermal conduction does not introduce
qualitatively new types of flow patterns in addition to those that
were discussed in \S3.1. However, it leads to
important quantitative differences. 
Firstly, the thermal conduction makes the contrasts of specific
entropy smaller, which leads to a suppression of 
the small-scale convection in the low  and moderate viscosity models.
Secondly, the thermal conduction acts as a cooling agent in the outflows,
reducing or even suppressing them in the models of moderate viscosity.
Calculations that include thermal conduction
cover a smaller region in the ($\alpha$, $\gamma$) plane
than those of the non-conductive models.
Figure~1 (right panel) summarizes some properties 
of the models with thermal conduction.
All computed models are stable. We have found only two types of flow patterns:
pure inflow (like in Model~E) for all models with $\alpha=0.3$, and
global circulation (like in Model~G) for models with smaller $\alpha$.

Figures~19 and 20 show the two-dimensional
structure of Models~N and P, respectively, in the
meridional cross section. Distributions of density and pressure are
almost spherical in both models. Important differences between the models
can be seen in the distribution of the momentum vectors (lower left panel)
and the Mach number (lower right panel).
In Model~N the momentum vectors are distributed almost spherically
at $r\la 500 r_g$. At $r\simeq 800 r_g$ in the polar directions there are
two stagnation points which divide inflows from outflows. The distribution
of the Mach number is flat, and the flow is significantly subsonic everywhere.
In Model~P the inward mass flux spreads in 
the wide ($\sim 45^\circ$) polar regions.
The equatorial inflow is relatively small. The distribution of the Mach
number in Model~P has an equatorial minimum and increases towards the  poles.
At $r\la 200 r_g$ the polar inflows are supersonic.

\begin{figure}
\plotone{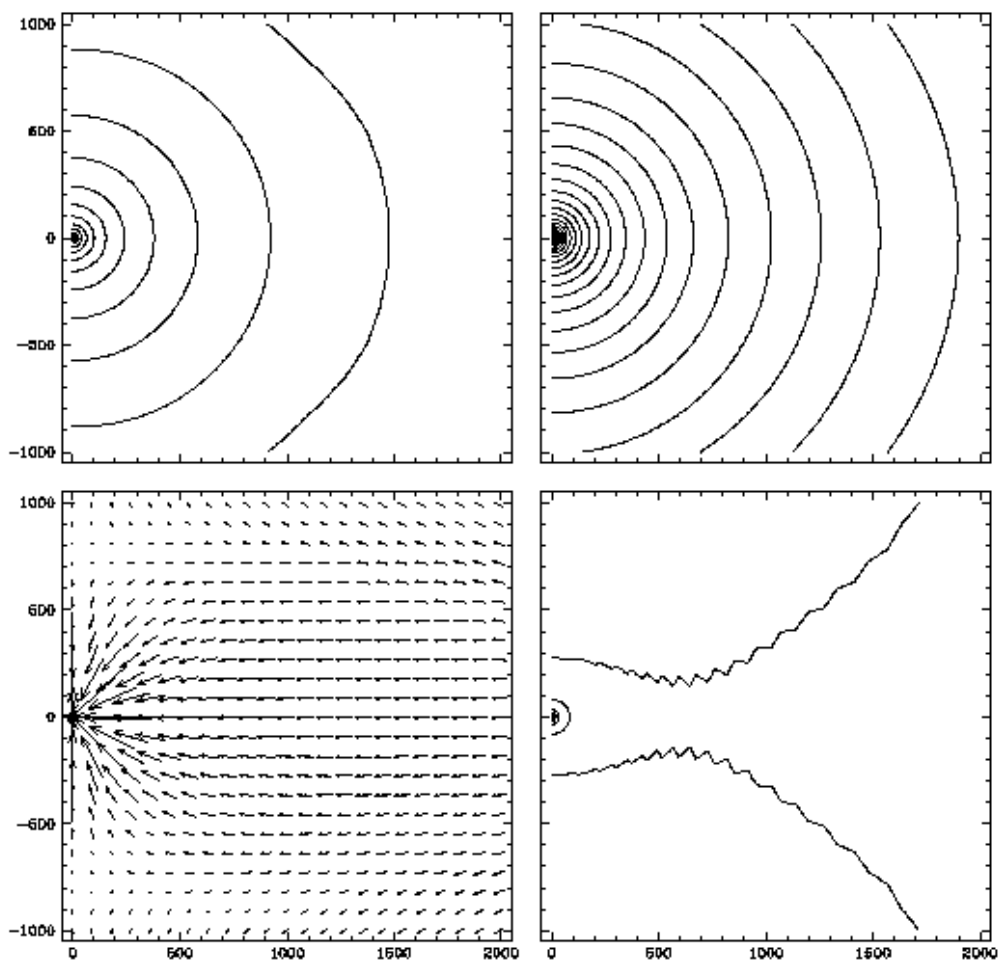}
\caption{Some selected properties of Model~N with thermal conduction
($\alpha=0.3$, $\gamma=5/3$ and $Pr=1$) 
in the meridional cross-section.
See the caption of Figure~3 for details.
The flow pattern is almost spherical at $r\la 500 r_g$.
There are two stagnation points at $r\simeq 800 r_g$, 
which divide the inflows and outflows in the polar directions 
(lower left panel).
\label{fig19}}
\end{figure}

\begin{figure}
\plotone{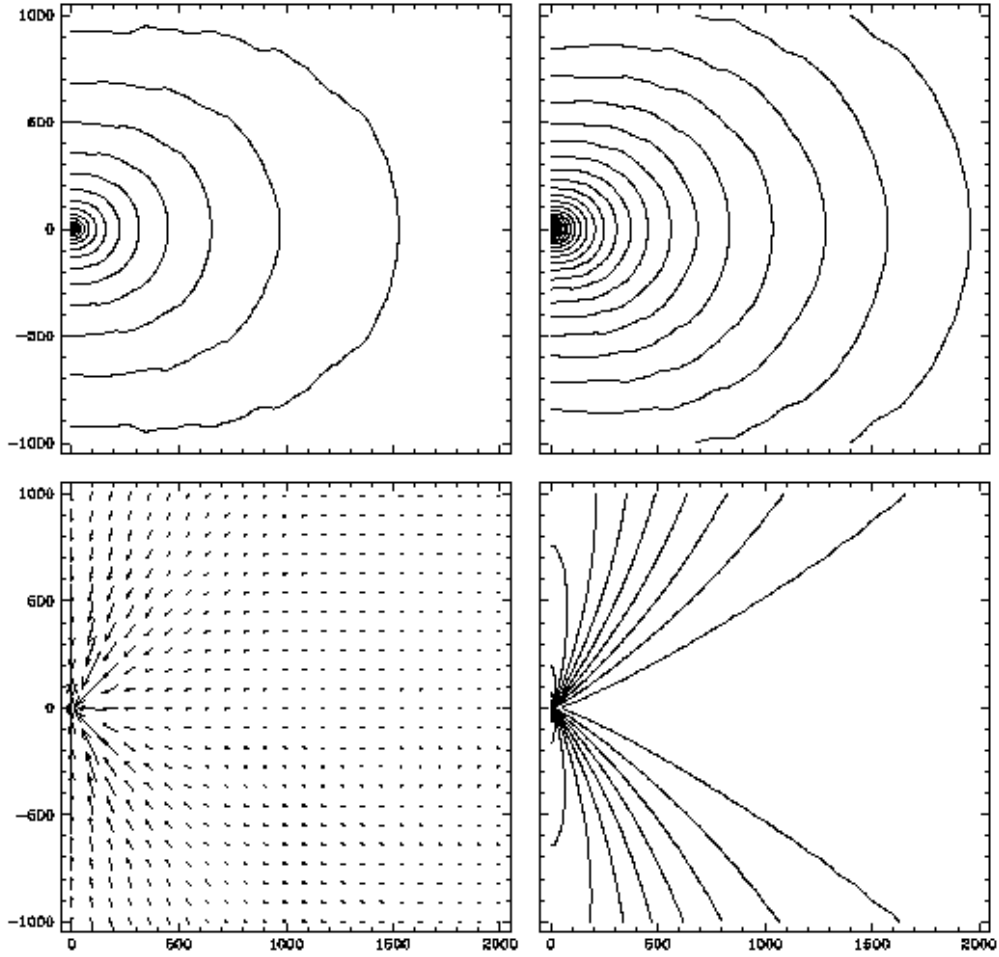}
\caption{Some selected properties of Model~P with thermal conduction
($\alpha=0.3$, $\gamma=4/3$ and $Pr=1$) 
in the meridional cross-section.
See the caption of Figure~3 for details.
The inward mass flux is concentrated in the polar directions, similar to 
the situation in Model~E (compare with Figure~10).
The thick line in the distribution of 
the Mach number (lower right) corresponds to ${\cal M}=1$.
\label{fig20}}
\end{figure}

\placefigure{fig19}
\placefigure{fig20}

Figure~21 presents the radial structure of the flow in Models~N (dashed lines), 
O (long-dashed lines) and P (dotted lines). The density profiles can
be approximated by a radial power-law, $\rho\propto r^{-\beta}$,
where the index $\beta\approx 1$ for Model~N, and $\beta\approx 1.5$ for
Models~O and P. The profiles of $\Omega(r)$ show a different behavior for each
model. In Model~N the values of $\Omega$ is significantly reduced with respect
to $\Omega_K$ in the inner region, $r\la 10^3 r_g$.
In Models~O and P the drop of $\Omega$ is less significant.
The ratio of $\Omega/\Omega_K$ goes to a limiting value in the latter
models: $\Omega/\Omega_K\simeq 0.6$ 
in the case of Model~P and $\simeq 0.3$ in the case of Model~O.
The radial inward velocities increase with decreasing $\gamma$
in the sequence of models from N to O and P, 
and is well described by the power-law
$v_r\propto r^{\beta-2}$. The profiles of $c_s(r)$ are weakly changed
from model to model.

\begin{figure}
\plotone{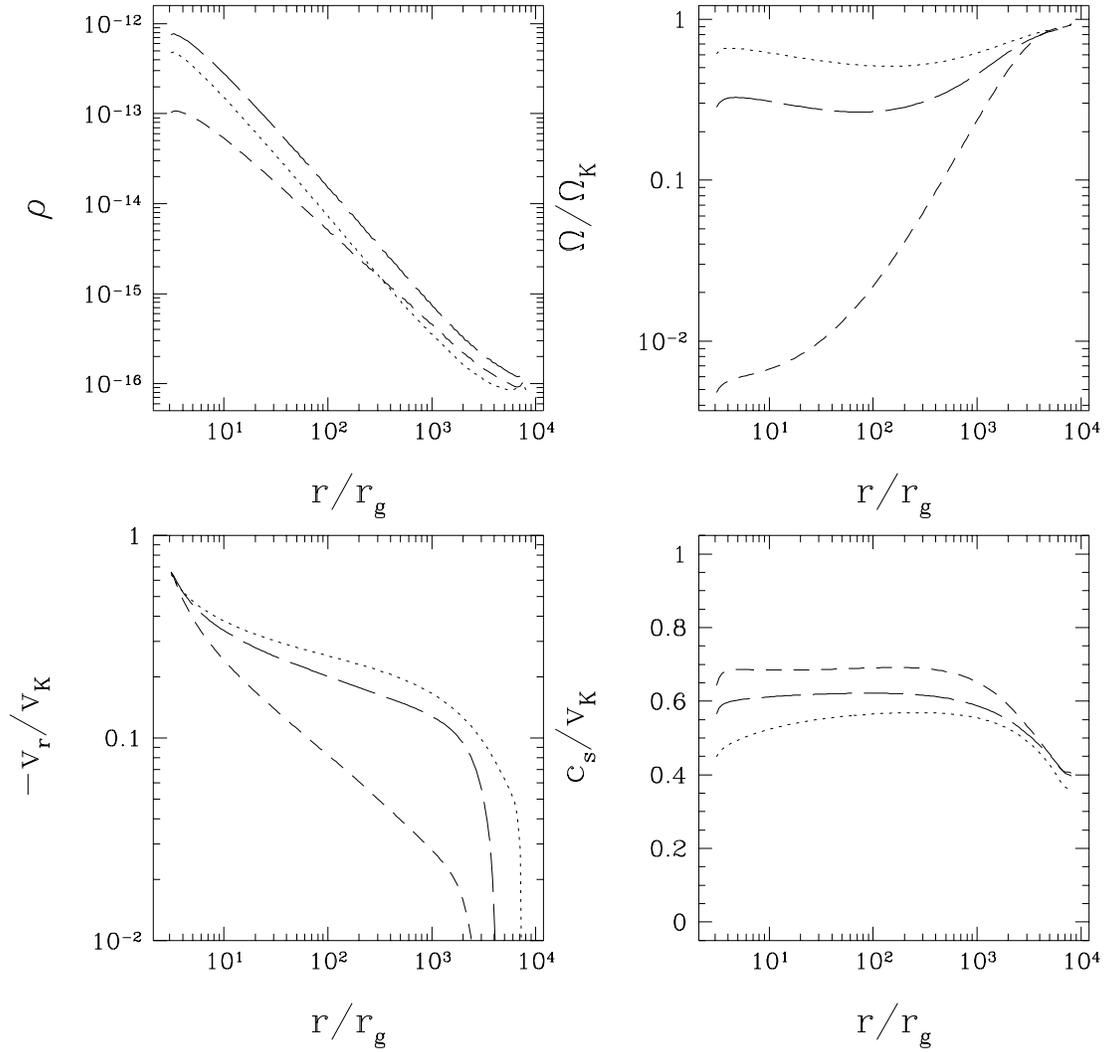}
\caption{Radial structure of the flow in Models~N (dashed lines),
O (long-dashed lines) and P (dotted lines).
All plotted quantities -- the density $\rho$,
angular velocity $\Omega$, radial velocity $v_r$
and sonic velocity $c_s$ -- have been averaged over the polar angle $\theta$.
\label{fig21}}
\end{figure}

\placefigure{fig21}

Models~Q, R and S have almost identical flow patterns and 
show weak dependence on $\alpha$ and $\gamma$. As an illustrative example
we present the flow pattern of Model~Q in Figure~22.
Like in the case of Model~G discussed in \S3.1.3
these models form stable global
circulations. Contrary to what is seen in Model~F, 
however, these models have a more pronounced
accretion funnel in the lower hemisphere, 
with a highly supersonic matter inflow.
Figure~23 shows the angular structure of Model~Q. The comparison 
of this models with
the properties of Model~G in Figure~13 does not indicate significant 
quantitative differences
in the flow structure except in the narrow polar regions. 

\begin{figure}
\plotone{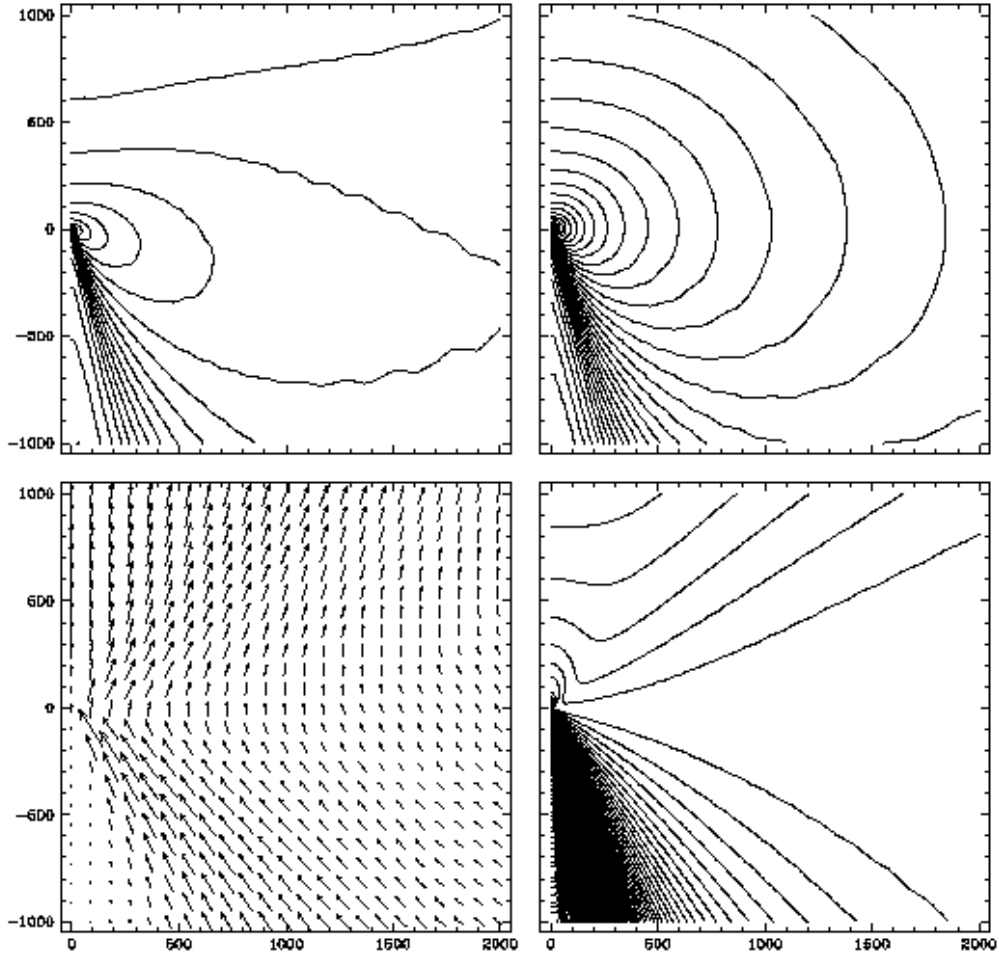}
\caption{Some selected properties of Model~Q with thermal conduction
($\alpha=0.1$, $\gamma=5/3$ and $Pr=1$)
in the meridional cross-section.
See the caption of Figure~3 for details.
The flow pattern is similar to that of Model~G (see Figure~12).
The thick line in the distribution of 
the Mach number (lower right) corresponds to
${\cal M}=1$.
\label{fig22}}
\end{figure}

\begin{figure}
\plotone{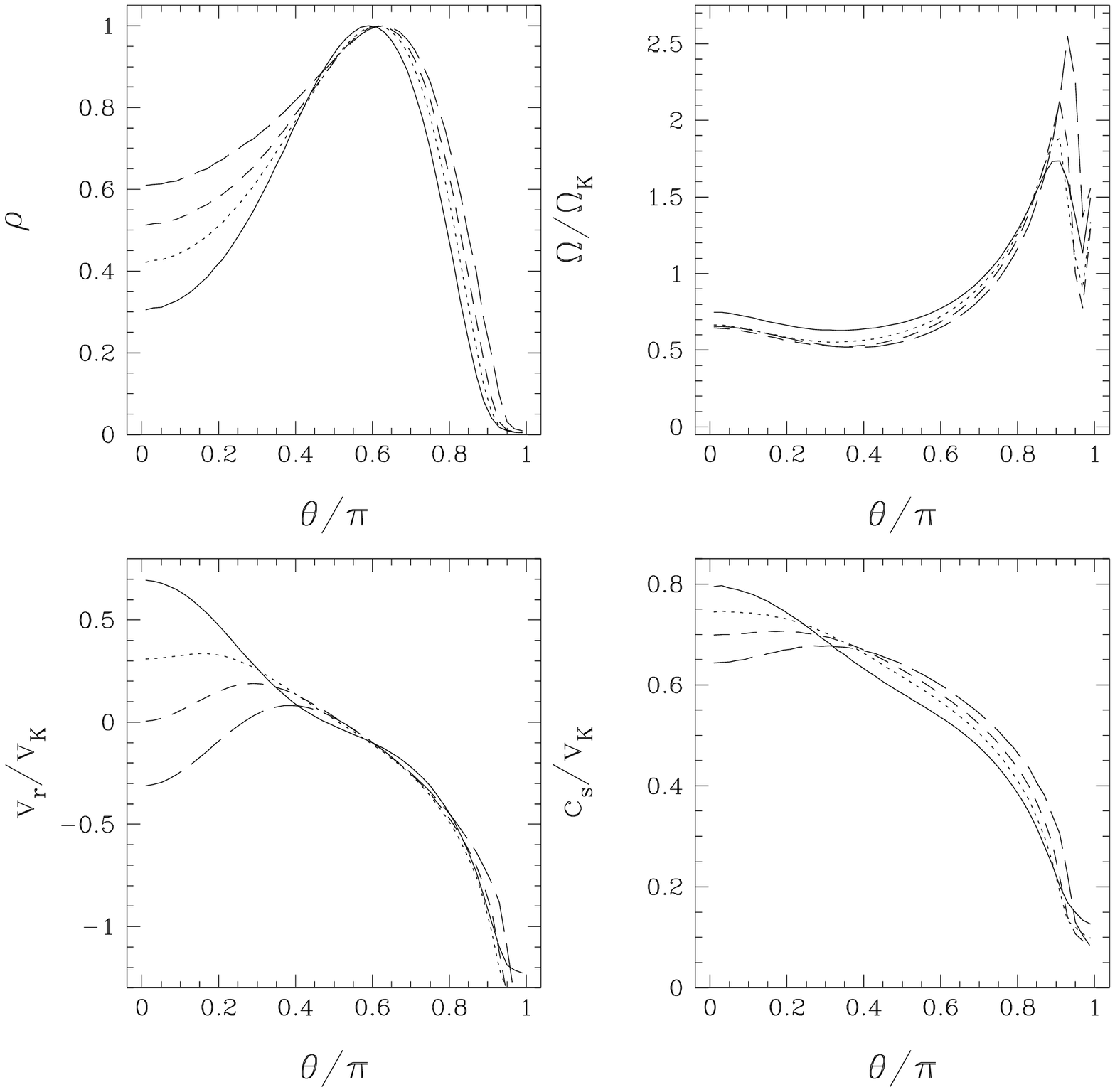}
\caption{Angular profiles of the density $\rho$, angular velocity $\Omega$,
radial velocity $v_r$ and sonic velocity $c_s$
from Model~Q  at four radial positions of
$r=30 r_g$ (long-dashed lines), $100 r_g$ (dashed lines), $300 r_g$
(dotted lines) and $1000 r_g$ (solid lines). The values of $\rho$ have been
normalized to the maximum value of $\rho$ at the corresponding radius.
\label{fig23}}
\end{figure}

\placefigure{fig22}
\placefigure{fig23}

\section{Discussion}

The properties of ADAFs are often discussed in terms of
one-dimensional (1D) vertically averaged analytic and numerical models.
The comparison of our two-dimensionl (2D) models and 
1D models of ADAFs constructed up to date
show important qualitative differences, which we would like to stress here.
Firstly, the 1D simulations of high and moderate viscosity 
($0.1\la\alpha\la 1$) flows cannot reproduce 
bipolar outflows and large-scale circulations due to obvious, intrinsic
limitations of the vertically averaged approach.
Only for the pure inflow models the 1D approach could be adequate.
However, ADAFs with pure inflows are realized only in a very small range of
the parameters $\alpha$ and $\gamma$ and therefore 
such a type of the flow is not general.
Secondly, the low viscosity ($\alpha\la 0.1$) 1D models of ADAF
constructed previously had not accounted at all, or had not accounted 
in all important details, the effects of convection. 2D models show that
the convection governs the structure of the low viscosity flows and 
significantly influences predicted observational properties of them.
Future low viscosity 1D ADAF models should account for the convection with
all the important details (see Narayan et al. 2000).

Convective accretion flows are
quite different from the `standard' self-similar ADAF solutions in several
points. Firstly, convective flows have a flattened density profile,
$\rho\propto r^{-\beta}$ with $\beta\simeq 0.5$--$0.7$ slightly
dependent on $\gamma$, whereas
$\beta=-3/2$ in the case of the self-similar ADAFs (Narayan \& Yi 1994). 
Secondly, there is a net outward
energy flux in these flows provided by convective motions, whereas 
the self-similar ADAFs have a zero net energy flux. 
The amount of energy transported outward by convections
is $\sim 10^{-2}\dot{M}_0 c^2$, 
and the value only moderately depends on 
$\alpha$ and $\gamma$. These effects have important implications
for the spectra and luminosities of accreting black holes.
Indeed, since $\rho\propto r^{-1/2}$ and $T\propto r^{-1}$, the
bremsstrahlung cooling rate per unit volume varies as
$Q_{br}\propto\rho^2 T^{1/2}\propto r^{-3/2}$ in the case of 
convective flows, and
$Q_{br}\propto r^{-7/2}$ in the case of self-similar ADAFs.
Integrating $Q_{br}$ over a spherical volume one can
demonstrate that most of the energy losses
in convective accretion flows occurs on the outside,
whereas most of the energy losses in self-similar 
ADAFs takes place in the innermost region.
Recent analytic works by Narayan et al. (2000) and
Quataert \& Gruzinov (2000) have provided considerable insight
into properties of accretion flows with convection. 
The analytic analysis is based on the ansatz
(confirmed by our previous and present numerical simulations) that
convective motions transport angular momentum inward, rather than outward.
%as it happens in the case of ordinary viscosity. 
All the basic properties
of numerical models with convection were reproduced 
in terms of a new self-similar solution in the mentioned works.
Note that the properties of convective models found in 2D simulations
have recently been confirmed in 3D simulations.
%(... 2000).

These properties of convective accretion flows 
could provide the physical explanation
of the phenomenon of `evaporation' of Shakura-Sunyaev thin disk 
with a following
formation of an ADAF in the innermost region: the convective outward
energy transport in optically thin ADAFs can power the evaporation process, 
in a similar way as that proposed by Honma (1996) 
in the case of the turbulent thermal conduction
(see, however, Abramowicz, Bj\"ornsson \& Igumenshchev 2000a for
a critical discussion of the Honma model).

In some respect the flows with large-scale circulations have a close
resemblance to those with convection.
All the basic properties of the two types of flows, including the flattened
radial density profiles and outward energy transport, are very similar.
A scenario in which the energy can be radiated by the gas
on the outside in flows with global meridional circulations
was discussed by Igumenshchev (2000).
It seems that in flows with large-scale circulations, the viscosity is large
enough to suppress the convection on scales $\la r$, but
convective motions with scale $\sim r$ still survive.

Self-similar ADAFs have always averaged $Be>0$, and it 
was argued that this can introduce the
formation of outflows (Narayan \& Yi 1994, 1995a).
Based on this idea, Blandford \& Begelman (1999) made the strong assertion
that {\it all} radiatively inefficient accretion flows must form
unbounded powerful
bipolar outflows (ADIOs, the 
advection-dominated inflows-outflows). Our previous and present investigations
do not confirm the physical consistency of the ADIOs idea.
Abramowicz, Lasota \& Igumenshchev (2000b) have stressed 
that, obviously, $Be>0$ is only
a necessary, but not a sufficient condition for unbounded outflows.
They provided an explicit numerical example in which a 2D ADAF with $Be>0$ 
has no outflows (see analogous Model~E in this study). 
Abramowicz et al. (2000b) have also
shown that $Be<0$ for low viscosity ADAFs that fulfill physically 
reasonable outer and inner boundary conditions, and have angular momentum
distribution close to that of Paczy\'nski's (1999) toy ADAF model.
Numerical results of Igumenshchev \& Abramowicz (1999) and of this paper
indicate that the convection energy transport provide an additional
cooling mechanism that always makes $Be<0$ in low viscosity ADAFs,
not only those with Paczy\'nski's-toy angular momentum distribution.
Thus, ADIOs do not exist if $\alpha\la 0.3$.
Models with very high viscosity, $\alpha\simeq 1$, do indeed show
behaviour similar to that postulated for ADIOs. However, as explained
in \S3.1.1, these models are significantly non-self-similar, and
therefore the Blandford \& Begelman's (1999) solution is not 
an adequate representation of them.

Thermal conduction was studied by Gruzinov (2000) in the context of
turbulent spherical accretion. He assumed the conduction
to be proportional to the temperature gradient, 
and found that the accretion rate
in this case can be significantly reduced compared with the Bondi rate.
Gruzinov's result is quite different to ours, because we found that
the conduction acts in such a way that the accretion rates increase.
These differences could indicate that the problem strongly depends
on the assumed prescription for thermal conduction.
% and/or on effects of geometry of flow.

\section{Conclusions}

We have performed a systematic study of 2D axisymmetric viscous rotating
accretion flows into black holes in which radiative losses are neglected.
Assumptions and numerical technique adopted here
are similar to those used by 
Igumenshchev \& Abramowicz (1999);
a few modifications are connected to inclusion of thermal conduction.
The thermal conduction flux was chosen to be proportional to the 
specific entropy gradient.

We assumed that mass is steadily injected within an equatorial torus near the
outer boundary of the spherical grid. 
The injected mass spreads due to the action of
viscous shear stress and accretes.
We set an absorbing inner boundary condition for the inflow at $r_{in}=3 r_g$.
We study the flow structure over three decades in radius 
using a variety of values of the viscosity parameter $\alpha$ and 
adiabatic index $\gamma$.

Our models without thermal conduction cover an extended region in the 
($\alpha$, $\gamma$) plane, $0.01\le\alpha\le 1$ and $4/3\le\gamma\le 5/3$.
We have found four types of pattern for the accretion flow, which had
been found earlier in two-dimensional simulations
by Igumenshchev \& Abramowicz (1999), Stone et al. (1999) and 
Igumenshchev (2000). The type of flow mainly depends on the value of $\alpha$,
and is less dependent on the value of $\gamma$. The high viscosity models,
$\alpha\simeq 1$, form powerful bipolar outflows. The pure inflow
and large-scale circulations patterns occur in the moderate viscosity models,
$\alpha\simeq 0.1-0.3$. The low viscosity ($\alpha\la 0.03$) models exhibit
strong convection. All models with bipolar outflows and pure inflow
are steady. The models with large-scale circulations could be either steady or
unsteady depending on the values of $\alpha$ and $\gamma$. 
All convective models are unsteady.

Some of our pure inflow and convective models do show a self-similar
behaviour. In particular,
the pure inflow model ($\alpha=0.3$, $\gamma=3/2$) reasonably well
satisfies the predictions of the self-similar solutions 
of Gilham (1981) and Narayan \& Yi (1994) in which $\rho\propto r^{-3/2}$. 
The convective accretion flows show a good agreement with
the self-similar solutions recently found by Narayan et al. (2000) and
Quataert \& Gruzinov (2000) in which $\rho\propto r^{-1/2}$.
The latter solutions have been constructed for the convection transporting
angular momentum towards the gravitational center.
The self-similar solutions for accretion flows with
bipolar outflows (ADIOs) proposed by 
Blandford \& Begelman (1999) have not been confirmed in our
numerical simulations.

The most interesting feature of the flows with large-scale circulations and
convective accretion flows is the non-zero outward energy flux, which
is equivalent to the effective luminosity $\sim 10^{-2}\dot{M}_0 c^2$, where 
$\dot{M}_0$ is the black hole accretion rate. 
This result has important implications for interpretation of
observations of accreting black hole candidates and neutron stars.

The accretion flows with thermal conduction have not been studied as completely
as the non-conductive flows. The conductive models show only two types
of laminar flow patterns: 
pure inflow (with $\alpha=0.3$) and global circulation
(with $\alpha\simeq 0.03-0.1$). The thermal conduction mainly acts as a cooling
agent in our models, it suppresses bipolar outflows and convective motions.

%\placefigure{fig2}

%\placetable{tbl-1}

%\acknowledgments

\noindent{\it Acknowledgments.}
The authors gratefully thank Ramesh Narayan for help with interpretation
of the numerical results and comments on a draft of the paper, 
Ed Spiegel for pointing out the importance of thermal 
conduction in viscous accretion flows, Rickard Jonsson for useful comments,
and Jim Stone and Eliot Quataert for discussions.
The work was supported by the Royal Swedish Academy of Sciences.

\clearpage

%   ----------- Table 1 ---------
%\colhead{$\Theta$\tablenotemark{b}}

\begin{deluxetable}{lccccc}
\footnotesize
\tablecaption{Parameters of the models. \label{tbl-1}}
\tablewidth{0pt}
\tablehead{
\colhead{Model} & \colhead{~~$\alpha$~~} & \colhead{~~$\gamma$~~} &
\colhead{~~${Pr}$~~} & \colhead{stability} &
\colhead{outflow(s)\tablenotemark{a}} 
}
\startdata
A & $1$    & $5/3$ & $\infty$ & stable & bipolar \nl
B & $1$    & $3/2$ & $\infty$ & stable & bipolar \nl
C & $1$    & $4/3$ & $\infty$ & stable & bipolar \nl
D & $0.3$  & $5/3$ & $\infty$ & stable & bipolar \nl
E & $0.3$  & $3/2$ & $\infty$ & stable & --- \nl
F & $0.3$  & $4/3$ & $\infty$ & unstable & unipolar \nl
G & $0.1$  & $5/3$ & $\infty$ & stable & unipolar \nl
H & $0.1$  & $3/2$ & $\infty$ & unstable & --- \nl
I & $0.1$  & $4/3$ & $\infty$ & unstable & --- \nl
J & $0.03$ & $5/3$ & $\infty$ & unstable & --- \nl
K & $0.03$ & $3/2$ & $\infty$ & unstable & --- \nl
L & $0.03$ & $4/3$ & $\infty$ & unstable & --- \nl
M & $0.01$ & $5/3$ & $\infty$ & unstable & --- \nl
N & $0.3$  & $5/3$ & $1$ & stable & --- \nl
O & $0.3$  & $3/2$ & $1$ & stable & --- \nl
P & $0.3$  & $4/3$ & $1$ & stable & --- \nl
Q & $0.1$  & $5/3$ & $1$ & stable & unipolar \nl
R & $0.1$  & $3/2$ & $1$ & stable & unipolar \nl
S & $0.03$ & $5/3$ & $1$ & stable & unipolar \nl
\enddata
\tablenotetext{a}{Only powerful outflows are indicated}
\end{deluxetable}

\clearpage

\begin{deluxetable}{lc}
\footnotesize
\tablecaption{The ratio of accretion rate to outflow rate. \label{tbl-2}}
\tablewidth{0pt}
\tablehead{
\colhead{Model~~~~~~} & 
\colhead{~~~~$\dot{M}_{0}/\dot{M}_{out}$\tablenotemark{a}~~~~} }
\startdata
A & $0.015$ \nl
C & $0.079$ \nl
D & $0.063$ \nl
E & $1.95$ \nl
F & $0.126$ \nl
G & $0.006$ \nl
I & $0.021$ \nl
J & $0.001$ \nl
K & $0.002$ \nl
L & $0.003$ \nl
\enddata
\tablenotetext{a}{$\dot{M}_{0}$ and $\dot{M}_{out}$ are the accretion and
outflow rates measured at the inner and outer numerical boundaries,
respectively.}
\end{deluxetable}

\clearpage

\clearpage

\clearpage

\clearpage

\clearpage

\clearpage

\clearpage

\clearpage

\clearpage

\clearpage

\clearpage

\clearpage

\clearpage

\clearpage

\clearpage

\clearpage

\clearpage

\clearpage

\clearpage

\clearpage

\clearpage

\clearpage

\clearpage

\end{document}